\documentclass[fleqn,usenatbib]{mnras}

\usepackage{newtxtext,newtxmath}

\usepackage[T1]{fontenc}
\usepackage{ae,aecompl}
\usepackage{graphicx}
\usepackage{booktabs}
\usepackage{multirow}
\usepackage{xspace}
\usepackage{comment}
\usepackage{siunitx}
\usepackage{overpic}
\DeclareSIUnit{\eV}{eV}
\DeclareSIUnit{\keV}{keV}
\DeclareSIUnit{\parsec}{pc}
\DeclareSIUnit{\erg}{erg}
\DeclareSIUnit{\gauss}{G}
\DeclareSIUnit{\year}{yr}
\usepackage{hyperref}
\newcommand{\eighteen}{RX\,J1856.5$-$3754\xspace}
\newcommand{\xmm}{\textit{XMM-Newton}\xspace}
\newcommand{\cha}{\textit{Chandra}\xspace}
\newcommand{\nicer}{\textit{NICER}\xspace}
\newcommand{\chisq}{\chi^2}
\newcommand{\sartore}{\citetalias{2012AA...541A..66S}}
\usepackage{natbib} 
\DeclareRobustCommand{\VAN}[3]{#2}
\newcommand{\flux}{\si{\erg\per\square\centi\meter\per\second}}
\title[Hard X-rays and spin down rate of \eighteen]{
Two decades of X-ray observations of the isolated neutron star \eighteen: detection of thermal and non-thermal hard X-rays and refined spin-down measurement
}

\author[De Grandis et al.]{Davide De Grandis$^{1}\thanks{E-mail: davide.degrandis@inaf.it}$, Michela Rigoselli$^{1}\thanks{E-mail: michela.rigoselli@inaf.it}$, Sandro Mereghetti$^{1}$, George Younes$^{2}$,
\newauthor
Pierre Pizzochero$^{3,4}$,
Roberto Taverna$^{5}$,
Andrea Tiengo$^{6,1,7}$,
Roberto Turolla$^{5,8}$
\newauthor
and Silvia Zane$^{8}$
\\
$^{1}$ INAF, Istituto di Astrofisica Spaziale e Fisica Cosmica Milano, via A.\ Corti 12, I-20133 Milano, Italy\\
$^{2}$ Department of Physics, The George Washington University, Washington, DC 20052, USA\\
$^{3}$ Dipartimento di Fisica, Università degli Studi di Milano, Via Celoria 16, 20133, Milano, Italy\\ $^{4}$ Istituto Nazionale di Fisica Nucleare, sezione di Milano, Via Celoria 16, 20133, Milano, Italy\\
$^{5}$ Dipartimento di Fisica e Astronomia, Università di Padova, Via F. Marzolo 8, I-35131 Padova, Italy\\
$^{6}$ Scuola Universitaria Superiore IUSS Pavia, Palazzo del Broletto, Piazza della Vittoria 15, 27100, Pavia, Italy\\
$^{7}$ Istituto Nazionale di Fisica Nucleare, Sezione di Pavia, Via A. Bassi 6, 27100, Pavia, Italy\\
$^{8}$ MSSL, University College London, Holmbury St. Mary, UK
}

\date{Accepted XXX. Received YYY; in original form ZZZ}

\pubyear{2022}

\begin{document}
\label{firstpage}
\pagerange{\pageref{firstpage}--\pageref{lastpage}}
\maketitle

%
%
 
  \begin{abstract}
The soft X-ray pulsar \eighteen is the brightest member of a small class of thermally-emitting, radio-silent, isolated neutron stars. Its X-ray spectrum is almost indistinguishable from a blackbody with $kT^\infty\approx\SI{60}{\eV}$, but evidence of harder emission above $\sim\SI{1}{\keV}$ has been recently found. 
We report on a spectral and timing analysis of \eighteen based on the large amount of data collected by \xmm in 2002--2022, complemented by a dense monitoring campaign carried out by \nicer in 2019.
Through a phase-coherent timing analysis we obtained an improved value of the spin-down rate $\dot{\nu}=\SI{-6.042(4)e-16}{\hertz\per\second}$, reducing by more than one order magnitude the uncertainty of the previous measurement, and yielding a characteristic spin-down field of \SI{1.47e13}{\gauss}.
We also detect two spectral components above $\sim$1\,\si{\keV}: a blackbody-like one with $kT^\infty=138\pm13\,$eV and emitting radius $31_{-16}^{+8}\,$m, and a power law with photon index $\Gamma=1.4_{-0.4}^{+0.5}$. 
The power-law 2--8\,keV flux, $(2.5_{-0.6}^{+0.7})\times\num{e-15}\si{\flux}$, corresponds to an efficiency of \num{e-3}, in line with that seen in other pulsars.
We also reveal a small difference between the 0.1--0.3\,\si{\keV} and 0.3--1.2\,\si{\keV} pulse profiles, as well as some evidence for a modulation above 1.2\,\si{\keV}. 
These results show that, notwithstanding its simple spectrum, \eighteen still has a non-trivial thermal surface distribution and features non-thermal emission as seen in other pulsars with higher spin-down power.

\end{abstract}


   \maketitle
%

\section{Introduction} 
Among isolated Neutron Stars (NSs), the so-called X-ray Dim Isolated NSs (XDINSs) represent a peculiar class of nearby sources, characterised by their thermal emission in the X-ray band, with faint optical counterparts and no confirmed detection of radio signatures \citep[e.g.][]{2007Ap&SS.308..191V,2009ASSL..357..141T}. Their very soft ($kT\lesssim \SI{100}{\eV}$) X-ray spectra are well reproduced by a simple blackbody with little interstellar absorption, with the additional presence of broad absorption lines in most sources \citep{2003A&A...403L..19H, 2004ApJ...608..432V,2004A&A...424..635H,2005ApJ...627..397Z}, and narrow, phase-variable ones in few cases \citep{2015ApJ...807L..20B,2017MNRAS.468.2975B}. These properties make XDINSs ideal study cases for models of NS thermal emission; this, in conjunction with their number, gained them the nickname of \emph{Magnificent Seven}.

In particular, \eighteen (in the following J1856 for short, \citealp{1996Natur.379..233W}) is the brightest ($f_X\simeq\SI{1.5e-11}{\erg\per\square\centi\meter\per\second}$) and closest ($d=123^{+11}_{-15}\,\si{\parsec}$, \citealp{2010ApJ...724..669W}) of the group, as well as the one showing the scantest amount of timing and spectral  features.
Its pulsation -- at $P=\SI{7.05}{\second}$ \citep{2007ApJ...657L.101T} and $\dot{P}=3\times\SI{e-14}{\second\per\second}$ \citep{2008ApJ...673L.163V} -- is hardly noticeable, due to a very small pulsed fraction $\rm{PF}\simeq1.2\%$.
Its X-ray spectrum resembles, to an excellent degree of accuracy, a pure blackbody emission with temperature $kT^\infty\approx\SI{60}{\eV}$ \citep{2001A&A...379L..35B, 2003A&A...399.1109B}, even though the emission in the optical band requires the presence of a complex emission mechanism and/or thermal structure \citep{1996ApJ...472L..33P, 2002ApJ...564..981P, 2007MNRAS.375..821H}. 
In fact, J1856 has been detected at optical wavelengths, too \citep[$V \sim 25.7$,][]{1997A&A...318L..43N}. The optical spectrum follows a $\lambda^{-4}$ law, but it lies above the extrapolation of a single-temperature X-ray blackbody at low energies. 

The brightness, simple spectrum, and steadiness of its emission make J1856 an ideal target for the calibration of X-ray telescopes \citep{2006A&A...458..541B}, and hence a very frequently observed object. In particular, \xmm  observed it about every six months, around April and October, since 2004 (with an earlier single observation in spring 2002). This large amount of data allows detailed spectral and timing studies.

Using all the 2002--2011 data from the \xmm EPIC-pn camera, \citet{2012AA...541A..66S}\defcitealias{2012AA...541A..66S}{Sar+12} (in the following \sartore) found that the X-ray spectrum of J1856 is indeed well described by a blackbody and set tight limits on possible spectral or flux variations. In addition, the high counting statistics allowed them to point out small instrumental systematic effects. In particular,  the derived best fit blackbody temperatures depend slightly on the position of the source on the detector. 
More recently, \citet{2017PASJ...69...50Y} and \citet{2020ApJ...904...42D} reported evidence for a flux enhancement above $\approx$1\,keV with respect to the blackbody model. This excess indicates a  spectral shape more complex than a single blackbody, possibly related to atmospheric effects and/or to the presence of a faint non-thermal emission component.

Here, thanks to the new wealth of data collected by \xmm in the last decade, we extend the work by \sartore\ and revisit the spectral and timing properties of J1856, using also data from the Neutron Star Interior Composition Explorer (\nicer) to derive an improved timing solution. The paper is organised as follows: in Sec.\ \ref{sec:data} we describe the data reduction; in
Sec.\ \ref{sec:spectrum} we present the spectral analysis, addressing in particular the systematics associated with the detector and the detection of a significant hard X-ray component; in Sec.\ \ref{sec:timing_sol} we present the coherent timing solution and pulse profiles resolved in different energy ranges; we discuss our findings in Sec. \ref{sec:discussion} and draw our conclusions in Sec.\ \ref{sec:conclusions}.

\section{Observations and data reduction}\label{sec:data}

This work is based on data collected between 2002 and 2022 by \xmm and in 2019 by \nicer. A log of all the observations is given in  Tables~\ref{tab:observations_xmm} and \ref{tab:observations_nicer}.

For  \xmm,  we used  data collected by the EPIC-pn camera,  reprocessed and analysed using the \xmm Science Analysis System (\textsc{SAS}) version 20.0.0 and the latest calibration files. 
The observations were conducted with the EPIC-pn camera operated in Prime Small Window mode, yielding a time resolution of \SI{5.7}{\milli\second}.
In order to exclude time intervals contaminated by a high background due to soft proton flares, for each observation we computed the distribution of high-energy count rates ($E>\SI{10}{\keV}$) binned at \SI{100}{\second}. The peaks of these distributions were then fitted with a Gaussian profile, and we discarded all the time intervals in which the count rate was more than $4\sigma$ apart from its mean. Thus, we obtained a net exposure time of 1.43 Ms of clean data.

For the spectral analysis we selected only EPIC-pn single-pixel events ($\textsc{pattern} = 0$). Most of the observations employed the thin optical filter; we discarded those taken with other filters, as well as the observation from Oct 2011 (\# 17 in the log), as it was split over four short intervals in which the source was in a CCD position quite far apart from those of the other observations (see the discussion in Sec.~\ref{sec:spectrum-X}). On the other hand, for the timing analysis, we used the whole set of observations and considered also double-pixel events (\textsc{pattern} $\leq$ 4) in order to maximise the counting statistics. 

For the timing analysis we also used a set of observations collected by \nicer. We reduced the data with the \textsc{nicerdas} software (version 9) including all the most recently released patches and calibration files (\textsc{caldb XTI20210720}). We filtered the data using \textsc{nicerl2} as part of heasoft version 6.30.1 and the standard cuts. 

We converted the times of all the \nicer and \xmm events to the solar system barycenter using the JPL ephemerides DE405 and the source nominal position R.A.=18$^{\rm h}$ 56$^{\rm m}$ 35$^{\rm s}$.795, Dec.=$\ang{-37;54;35}\!.4$ \citep{2010ApJ...724..669W}. 

\section{Spectral analysis} \label{sec:spectrum}

The X-ray emission from J1856 is extremely soft, with the great majority of the photons detected below $\approx$1\,\si{\keV}. The counting statistics in this range is so large that subtle instrumental effects and calibration uncertainties, which are generally neglected for fainter sources, are clearly visible (see Section~\ref{sec:spectrum-X}). 

On the other hand, the source is very faint at higher energies, where it can be detected only by summing many \xmm observations. Therefore, for the analysis at energies above 1.2\,\si{\keV} we used a maximum-likelihood (ML) method as described in \citet[][and references therein]{2021A&A...646A.117R}. For faint sources, this method is more effective than standard spectral analysis because it fully exploits all the counts distributed according to the instrumental point spread function (PSF) and it estimates the background locally, instead than from separate regions of the CCD. In the whole ML analysis we used both single- and double-pixel events.

\begin{figure}
    \centering
    \includegraphics[width=.49\textwidth]{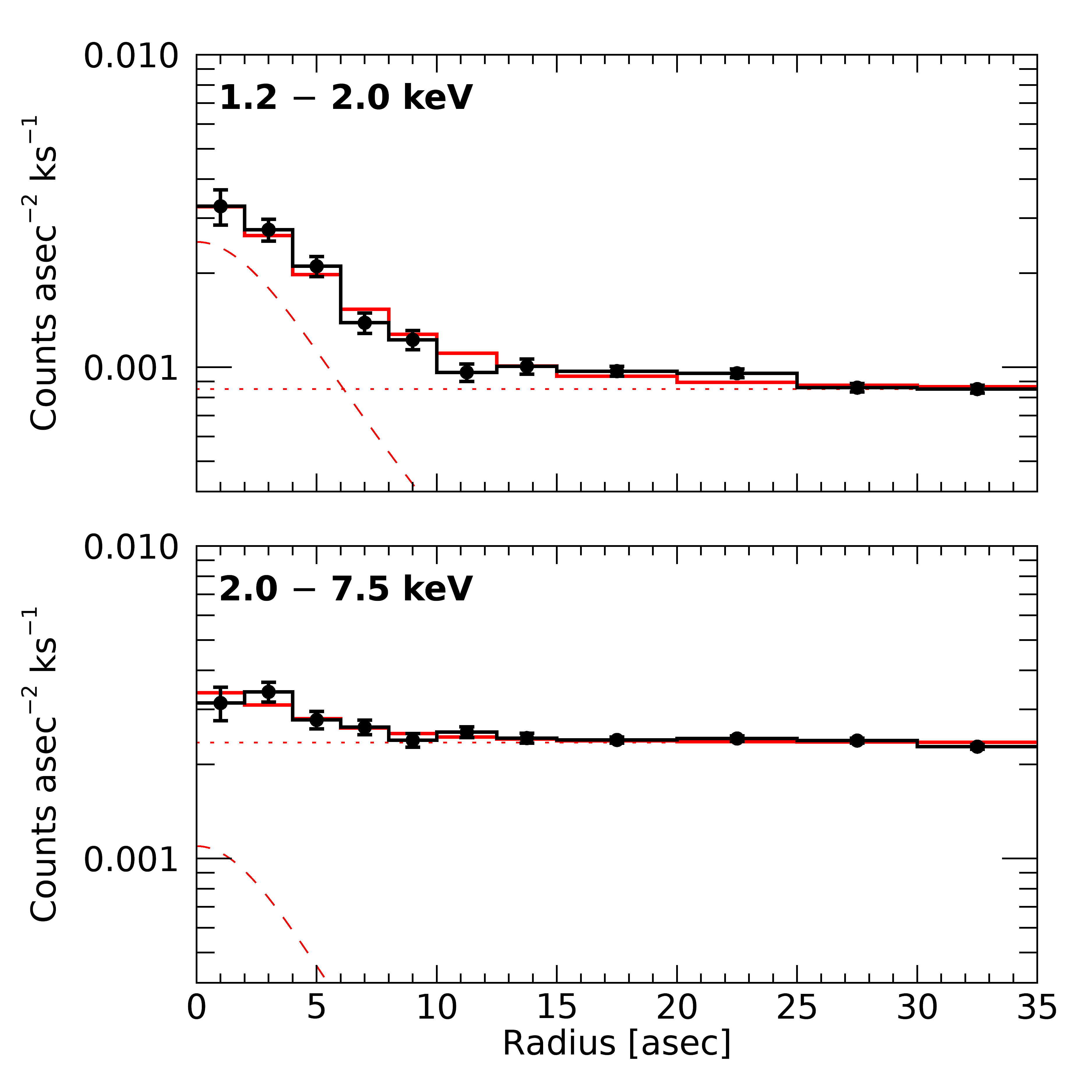}
    \caption{Surface brightness radial profiles of J1856 
    in the 1.2--2\,\si{\keV} (upper panel) and 2--7.5\,\si{\keV} (lower panel) energy bands. The red solid lines are the expected profiles, given by the sum of the instrument PSF (dashed) and a spatially uniform background (dotted).}
    \label{fig:radprofile}
\end{figure}

Based on the instrumental positions of the source in each observation, as measured in the 0.4--0.5\,\si{\keV} range where the signal to noise is highest, we realigned all the observations to correct for small pointing differences, and then stacked them to produce summed images in the 1.2--2\,\si{\keV} and 2--7.5\,\si{\keV} ranges. 
Applying the ML source detection to these images, we obtained $817\pm57$ and $327\pm64$ net source counts in the two energy ranges, corresponding to a detection significance of 18.2$\sigma$ and 5.5$\sigma$, respectively. 
Fig.\ \ref{fig:radprofile} shows the radial profiles of J1856 between 1.2--2\,\si{\keV} (upper panel) and  2--7.5\,\si{\keV} (lower panel), compared to those expected for a point source in these energy ranges (red lines, see \citealt{simo}).
In order to extract the spectra and pulse profiles at $E>1.2$\,\si{\keV} described below, we applied the same ML methods to images accumulated in different energy and/or phase intervals.

\subsection{Long-term behaviour in soft X-rays} \label{sec:spectrum-X}

We first investigated the possible long term evolution of J1856 by a spectral analysis of its 
0.16--1.2\,\si{\keV} flux (the same energy range explored in \sartore) in the individual \xmm observations.  
The source counts were extracted from  circular regions with  $\SI{30}{\arcsecond}$ radius and
the background counts from  circles of radius $\SI{40}{\arcsecond}$ located as far from the source as possible, within the Small window field of view. The spectra were then rebinned to have at least $25$ counts per bin.

\begin{figure*}
\begin{overpic}[percent,width=\textwidth]{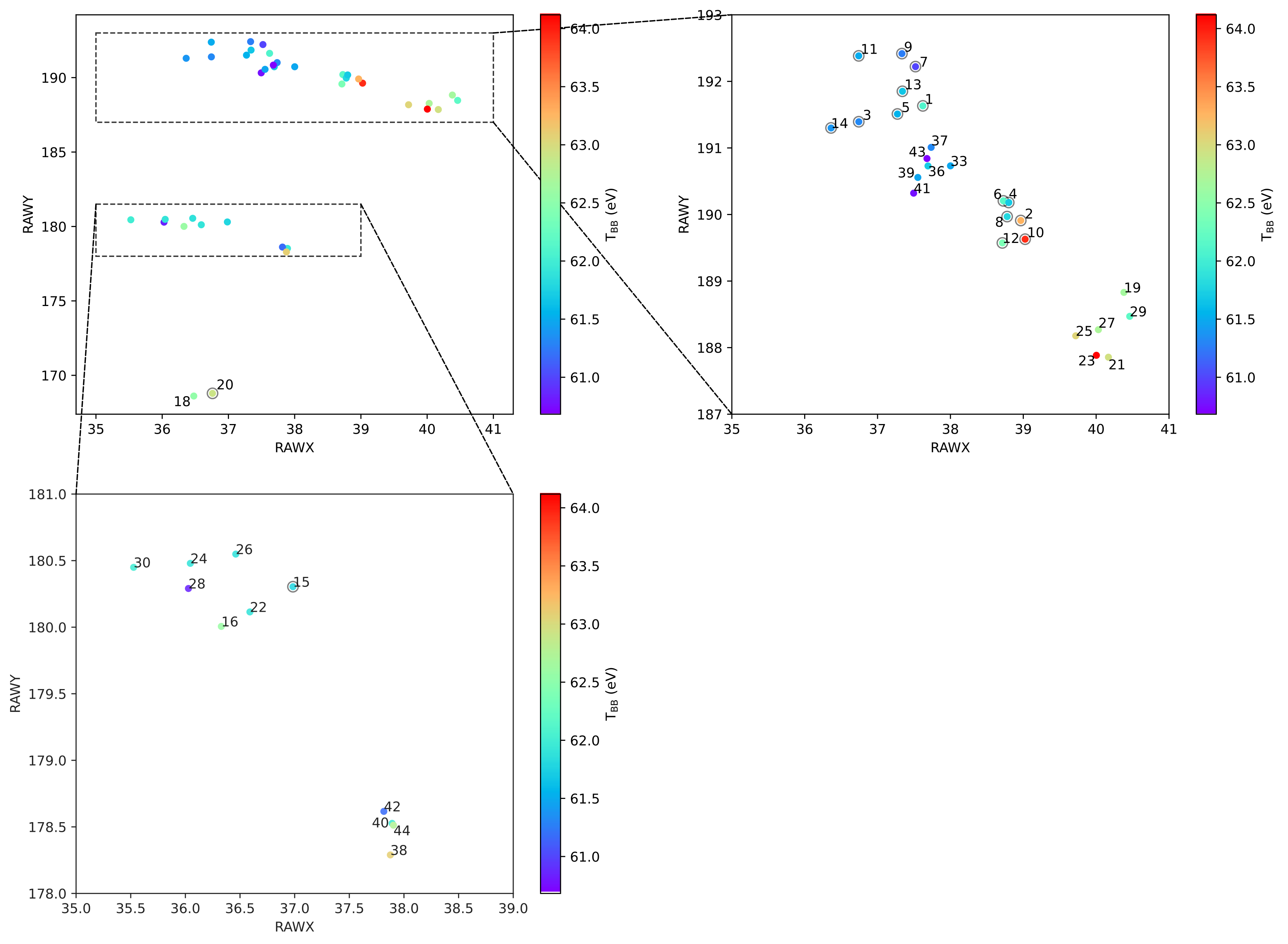}
\put(50,0){\includegraphics[width=.5\textwidth]{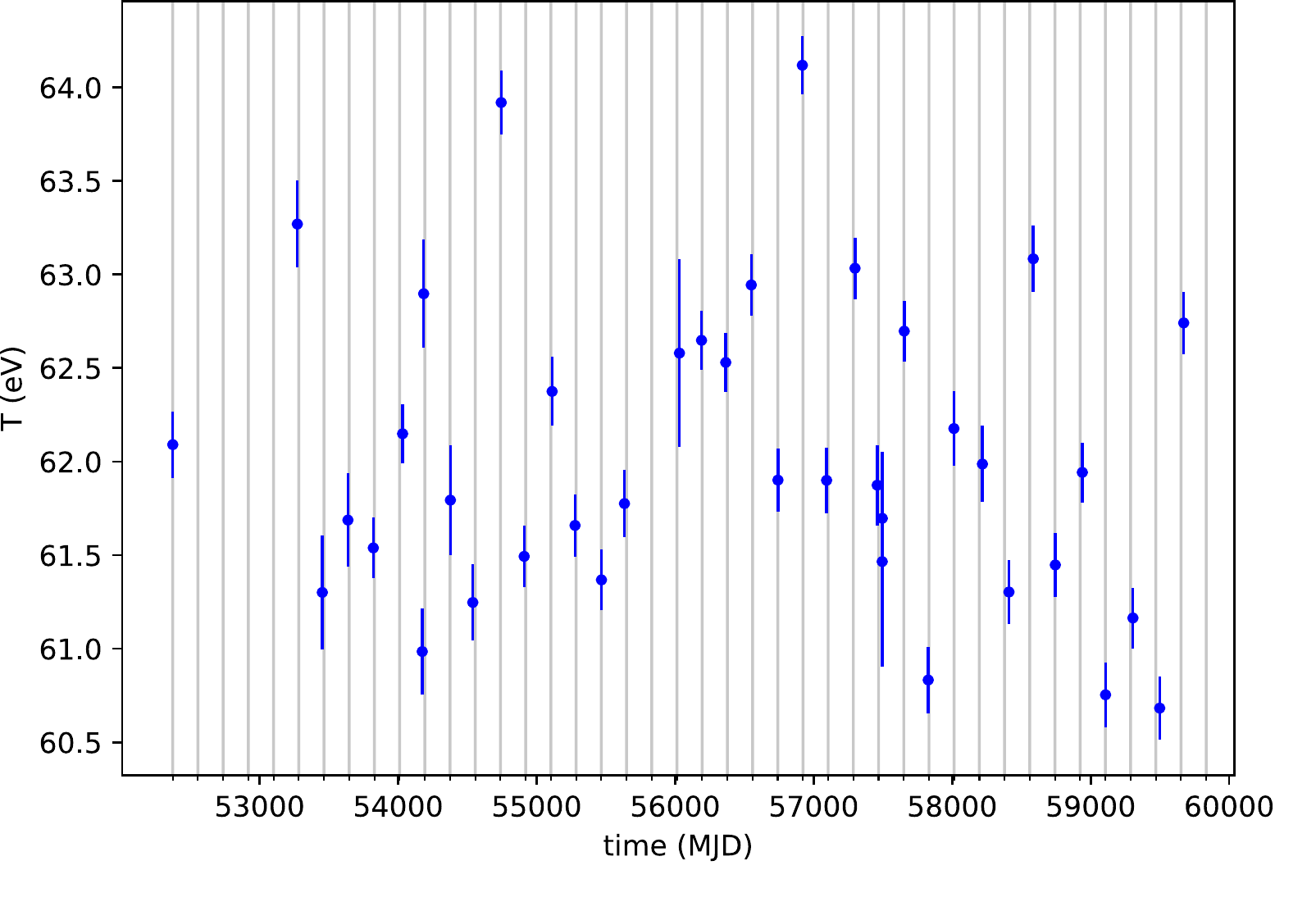}}
\end{overpic}
\caption{\emph{Top-left panel}: Position of the source on the detector in raw coordinates
(the top-right and bottom-left panels are zoomed views of the regions in the insets). The colours indicate the temperature obtained from the single blackbody fit and the labels correspond to the observation numbers of Table~\ref{tab:observations_xmm}. The circled points are the observations already analysed by \sartore; note, in particular, the two clusters they form in the top-right panel. \emph{Bottom-right panel}: Temperatures obtained from the single blackbody fits as a function of time. The vertical lines are spaced by 6 months to highlight the seasonal cadence of the observations.}
    \label{fig:detector}
\end{figure*}

We performed a simultaneous fit of the individual spectra with an absorbed blackbody model, letting the three fit parameters for each observation free: we found hydrogen column densities in the range $N_{\rm H}$ = [$2-11$]$\,\times10^{19}$ \si{\per\square\centi\meter}, and observed temperatures in the range $kT^{\infty}$ = [$60.7-64.1$] \si{\eV} ($\chisq=6421.00\text{ with }5421\text{ dof}$); these values are shown in Fig.~\ref{fig:detector} as a function of time and position of the source on the detector. We repeated the fit with all the column densities linked to a common value, finding $N_{\rm H}=(3.95\pm0.08)\times\SI{e19}{\per\square\centi\meter}$; the blackbody temperatures and normalisations show an analogous pattern and are within $2\sigma$ from the values of the previous fit ($\chisq=6689.30\text{ with }5460\,\text{dof}$).

Even though the derived temperatures are in a narrow interval, they are not compatible with each other if only the statistical errors of the fits are considered. No discernible trend of temperature as a function of time and/or season of the observation is present. With a similar analysis of the data taken before 2010, \sartore\ noted that the temperature depends on 
the position of the source on the CCD (encoded by the RAWX and RAWY coordinates). We found similar results, but the new data acquired after 2011 indicate a more complicated pattern of position-dependent values than that shown by the older data.
The lack of a clear pattern of this systematic effect makes it impossible
to select a group of observations more reliable or well behaved than the others, as it was done by \sartore.
Although this prevents a precise estimate of the intrinsic thermal evolution of J1856, we are able to rule out a variation of more than few \si{\eV} over the $\approx\SI{20}{\year}$ time period we considered.

\begin{figure}
    \centering
    \includegraphics[width=.48\textwidth]{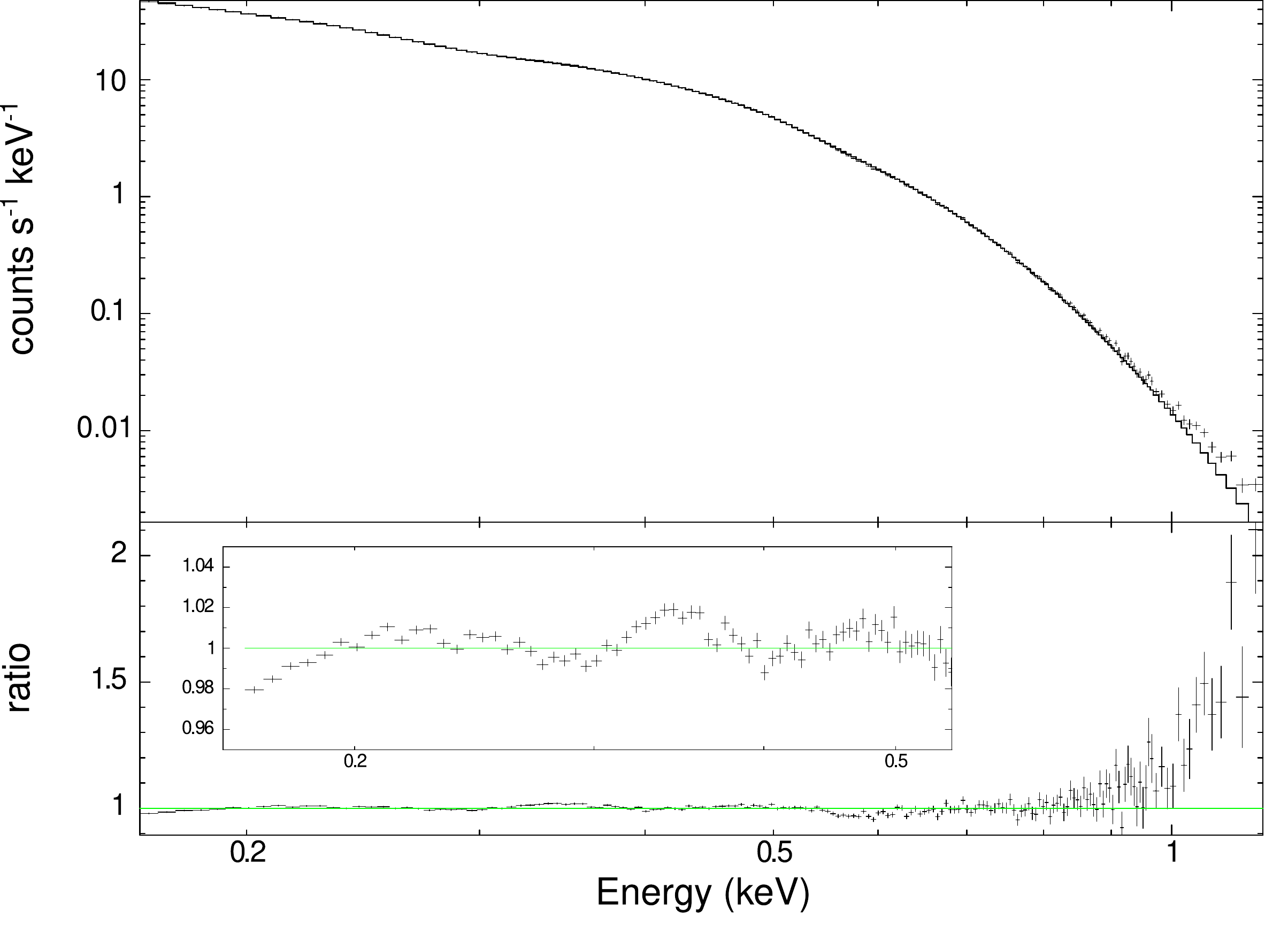}
    \caption{Total (summed) spectrum of all the data in the [$0.16$--$1.2$]\,\si{\keV} band fitted with single blackbody. The bottom panel shows the ratio between the data and the folded model, highlighting the presence of a hard excess. The zoomed in version of the ratio below $\SI{0.6}{\keV}$ in the inset shows that the residuals are at a level $\lesssim2\%$, well below the instrument effective area uncertainties.}
    \label{fig:1bb_sum}
\end{figure}

Given these instrumental effects affecting the derived parameters, in the following we will analyse the total spectrum (sum of all the spectra from the single observations) introducing a systematic error of $1\%$ in the spectral fits. The best fit with an absorbed blackbody to the total spectrum, binned with a minimum of 200 counts per bin, is shown in Fig.\ \ref{fig:1bb_sum}; it gives $kT^\infty=62.51\pm0.05\,\si{\eV}$ and normalisation corresponding to an observed emission radius of $R^\infty=4.74\pm0.02\,\si{\kilo\meter}$ ($\chisq=318.38$ with 169 dof). 
The fit residuals indicate the presence of a hardening above $\sim$1\,\si{\keV}, which was undetectable in the spectra of the single observations and will be investigated in the next subsection.
Note that the residuals at $E<0.6$\,\si{\keV}, highlighted in the figure inset, are at the level of only $\lesssim2\%$, well below the current uncertainty of the instrument effective area calibration \citep[e.g.][]{2014A&A...564A..75R}. Furthermore, the width of these deviations is narrower than the EPIC-pn energy resolution at these energies. Therefore, they do not imply the presence of spectral features in the X-ray emission of J1856.   

\subsection{X-ray spectrum}
\begin{figure}
    \centering
    \includegraphics[width=.49\textwidth]{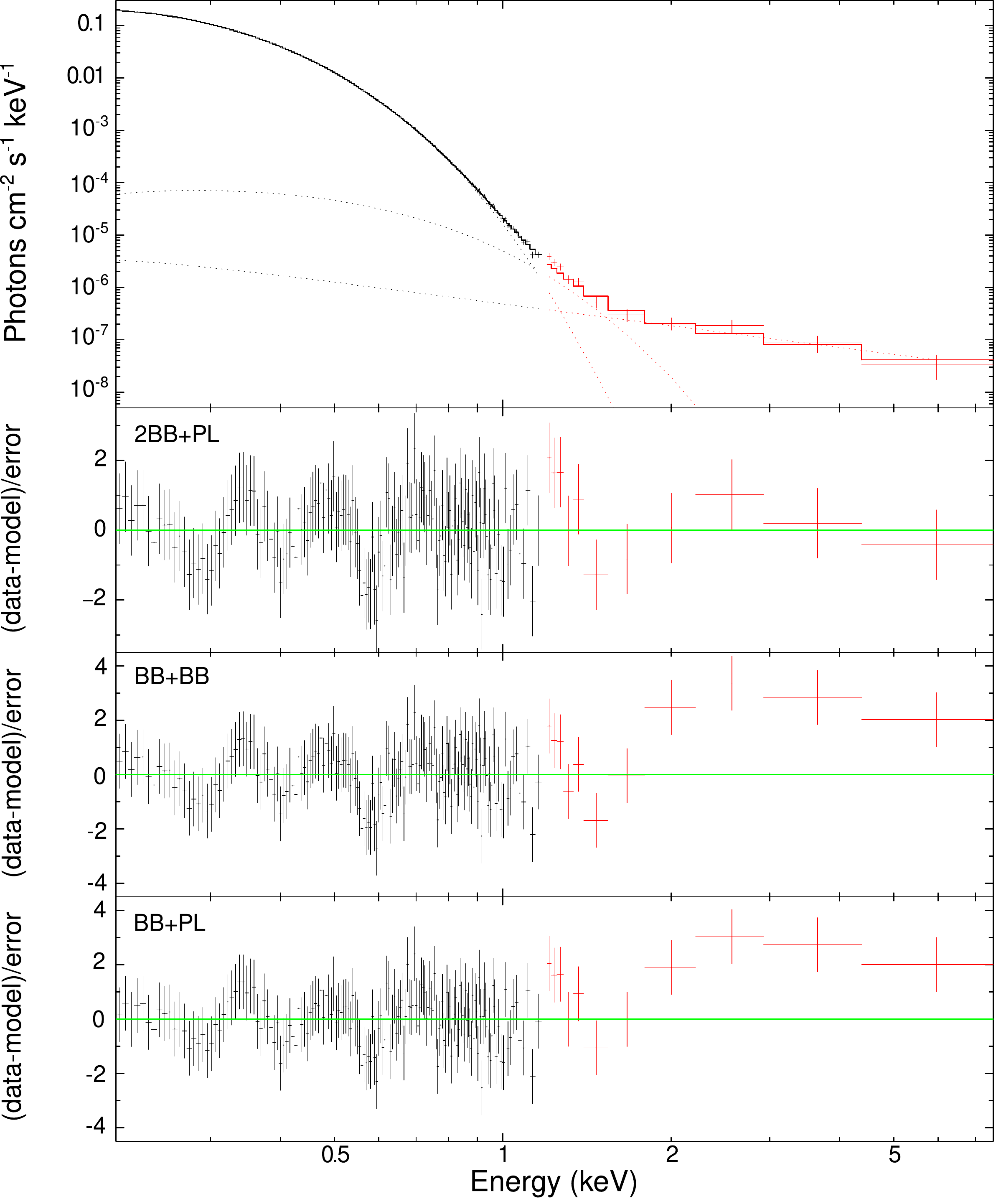}
    \caption{Top panel: Total X-ray spectrum obtained with the traditional background subtraction method (soft part, in black) and the ML method (hard part, in red). Out best-fitting model with two blackbody and a power-law components is superimposed in the top panel. The three bottom panels show the residuals for the two-blackbody plus power law fit (2BB+PL) itself, for the two blackbody (2BB) fit, and for the blackbody plus power law (BB+PL) fit (note the different vertical scale in the latter cases). We reject the two latter models which yield unsatisfactory residuals at $E>1$\,\si{\keV}.}
    \label{fig:spec_all}
\end{figure}

Given the statistically significant detection of J1856 also above 1.2\,\si{\keV}, we now consider the spectrum in the whole 0.2--7.5\,\si{\keV} range.
In order to fit the excess with respect to the single blackbody, we first tested the addition of a second component, either a blackbody or a power law.
Both models gave formally acceptable fits 
($\chisq=184.28$ and $\chisq=167.89$ with $171$ dof, respectively)
because the $\chisq$ values are dominated by the good matching of the blackbody in the low energy range. However, the residuals indicate that these models are not a good description of the spectra above $\sim$1\,\si{\keV} (see the two bottom panels of Fig.\ \ref{fig:spec_all}).

An acceptable fit is obtained with a three-component model, namely the sum of two blackbodies and a power law ($\chisq=156.10$ with $169$ dof, see the best-fitting parameters in Table~\ref{tab:hard_components} and Fig.~\ref{fig:spec_all}). 
As explained above, the wiggles in the residuals below $\sim$0.6\,\si{\keV} are of instrumental origin. 
Alongside a blackbody component akin to the one found so far ($kT_1^\infty = 61.9 \pm 0.1$\,eV, $R_1^\infty = 4.92_{-0.06}^{+0.04}$\,km) that describes the softest part of the spectrum, a second, hotter and from a smaller emission area ($kT_2^\infty = 138 \pm 13$\,eV, $R_2^\infty = 31_{-16}^{+8}$\,m) 
is found. This accounts for most of the emission observed around $\approx$1.5\,\si{\keV}. At even higher energies, the emission is dominated by a power law with photon index $\Gamma = 1.4_{-0.4}^{+0.5}$, and flux in the $2$--$\SI{8}{\keV}$ band of $(2.5_{-0.6}^{+0.7})\times 10^{-15}$ \flux.
Conversely, a model with a single blackbody and two power laws 
($\chisq=144.89$ with $169$ dof), yields an unphysically large photon index ($\Gamma\sim6$) for the additional power-law component.
   
\begin{table}
    \centering
    \caption{Best fit of the   X-ray spectrum.}
    \label{tab:hard_components}
    \begin{tabular}{clr}
         \toprule
         \midrule
         Component & Quantity & Value \\\midrule
         & $N_{\rm H}$ $(\SI{e19}{\per\square\centi\meter})$ \dotfill& $3.9_{-0.3}^{+0.3}$ \\\addlinespace[0.8em]
         \multirow{2}*{Soft BB}& $kT^\infty$ (\si{\eV})  \dotfill &$61.9_{-0.1}^{+0.1}$ \\\addlinespace[0.4em]
                                    & $R^\infty$ (km)    \dotfill  & $4.92_{-0.06}^{+0.04}$  \\\addlinespace[0.8em] 
                                    
         \multirow{2}*{Hard BB}& $kT^\infty$ (\si{\eV})\dotfill&$138^{+13}_{-13}$ \\\addlinespace[0.4em]
                                    & $R^\infty$ (m)     \dotfill     & $31^{+8}_{-16}$  \\\addlinespace[0.8em] 
         \multirow{2}*{PL tail}& 
         $\Gamma$\dotfill& $1.4^{+0.5}_{-0.4}$ \\\addlinespace[0.4em] 
         & $I_\text{[2--8]}$ (\SI{e-15}{\erg\per\square\centi\meter\per\second})\dotfill& $2.5_{-0.6}^{+0.7}$
         \\
         \midrule
         & systematic\dotfill & 1\%\\
         & $\chisq$ / dof \dotfill& $156.10/169$ \\
         & nhp \dotfill& $0.75$\\
         \bottomrule\\[-5pt]
    \end{tabular}

\raggedright
Radii are computed taking $d = \SI{123}{\parsec}$ \citep{2011MNRAS.417..617T}; $I_\text{[2--8]}$ indicates the flux in the 2--8\,\si{\keV} band.
\end{table}

\section{Timing analysis} \label{sec:timing_sol}

\subsection{Coherent timing solution}
\label{sec:coher_ts}

\begin{figure*}
    \centering
    \includegraphics[width=.75\textwidth]{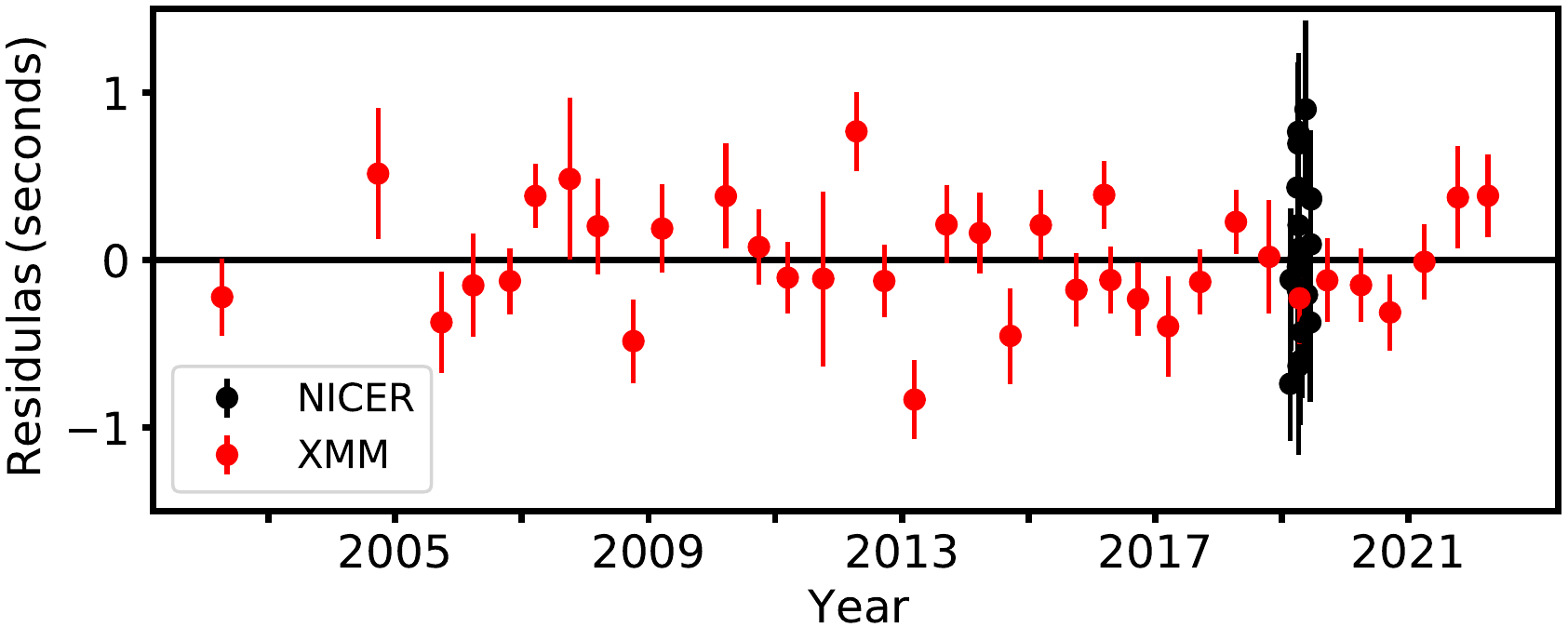}
    \caption{Residuals of the phase-connected timing solutions with $\nu=0.14173907778(8)$ Hz and $\dot{\nu}=-6.042(4)\times 10^{-16}$ Hz s$^{-1}$.}
    \label{fig:timing}
\end{figure*}

We start our timing analysis of J1856 by utilising the \nicer observations, which were taken with heavy cadence and long exposure in the period from 2019 April 01 until 2019 April 07. During this time span and with an exposure totalling 60~ks, we collect $\approx6\times10^5$ counts in the energy range 0.2--1\,\si{\keV}. We apply a $Z^2_{1}$ search, i.e., Rayleigh test \citep{1983A&A...128..245B}, 
around the expected source spin frequency, $0.1417\pm0.0001$~Hz. We detect a strong signal at $\nu=0.1417392(1)$~Hz at a confidence level $>99.99\%$ at epoch $t_0=58576.5$~MJD (TDB).

We refine and update this initial measurement through a phase-coherent timing analysis technique \citep[e.g.][]{2003ApJ...599..485D}. Using the high signal-to-noise pulse profile corresponding to the above frequency, we derive a model of the source pulse shape in the 0.2 to 1\,\si{\keV} band, consisting in the fundamental of a Fourier series. Then, we divided the unbinned data in chunks comprising $10^5$~counts and collected in a time-span not exceeding 7 days. This number of counts ensures the detection of the pulse at a $\sim3\sigma$ level.
We fit each unbinned light curve with the pulse model with the maximum likelihood estimate method and only allowing for a phase shift $\Delta\phi$ \citep[see][for more details]{2009ApJ...706.1163L,2011ApJS..194...17R}. Finally, we fit these phase shifts to a polynomial of the form $\delta\phi = (t-t_0)\nu + 1/2(t-t_0)^2\dot\nu + 1/6(t-t_0)^3\ddot\nu + \ldots$, truncated at the highest significant term according to an F-test.

Between 2019 February and June, we derive 19 pulse times of arrival (ToAs) utilising the \nicer data. We can describe these pulse ToAs with a simple correction to the source spin frequency. 
This resulted in $\nu=0.141739071(4)$~Hz at the same $t_0$ given above. A frequency derivative in this time span is not statistically required and we derive a $3\sigma$ upper limit of
$|\dot\nu| < 3.0\times10^{-15}$~Hz~s$^{-1}$.

The small uncertainty on the spin frequency of $4\times10^{-9}$~Hz allows us to extend our phase-coherence up to $\pm1$~year around $t_0$ without the loss of cycle count, even when $\dot\nu$ is close to the $3\sigma$ upper-limit value. Hence, we derive the ToAs from the \xmm data that are within this time span. We follow the same technique to compute the ToAs, yet, we used each full \xmm observation to derive one ToA given the relatively short exposure of each of them, based on data in the 0.1--1\,\si{\keV} range. With a baseline of 700~days, we observe a clear curvature in the pulse ToAs which can be well described by introducing a second term to the timing model, $\dot\nu=-6.1(4)\times10^{-16}$~Hz~s$^{-1}$.

Following the above methodology, we iteratively added more ToAs while refining our timing solution, ensuring that no cycle ambiguity exists at each step. We find that our full baseline, spanning about 20 years, can be well described by a timing model including only $\nu$ and $\dot\nu$. This model results in a $\chisq$ of 83.4 for 52 dof and a root mean square (rms) residual of 0.054~cycle. The full timing model is presented in Table~\ref{tab:timing}, while the residuals of the pulse ToAs in seconds are shown in Fig.~\ref{fig:timing}.
We, moreover, find a hint for a second derivative of the frequency $\ddot{\nu} = -1.2(7) \times 10^{-26}$ Hz s$^{-2}$, but the improvement of the best fit ($\chisq = 80.3$ for 51 dof) is marginal (F-test probability of 0.167).

We checked that our results are not significantly affected by the proper motion of J1856 ($\mu_\alpha=+325.86\pm 0.21$ and $\mu_\delta=-59.22\pm 0.18$ mas~yr$^{-1}$, \citealt{2010ApJ...724..669W}). 
In fact, by  applying the barycentric corrections with the different source positions in each observations, the derived
ToAs vary by less than 0.011 s, that is three times smaller than each ToA rms ($\approx$0.381~s). 

\begin{table}
\centering
\caption{Parameters of the timing solution.}
\label{tab:timing}
\begin{tabular}{lc}
\toprule
\midrule
Quantity\hspace{2.5cm} &Value \\
\midrule
MJD $t_0$ (TDB)\dotfill & 58$\,$576.5 \\
MJD range\dotfill & 52$\,$372 -- 59$\,$673 \\
$\nu$ (Hz)\dotfill      &  0.14173907778(8) \\ 
$\dot{\nu}$  (Hz s$^{-1}$)\dotfill   & $-6.042(4) \times 10^{-16}$  \\ 

$\chisq$/ dof\dotfill    &   83.4 / 52 \\
ToA rms (s)\dotfill & 0.381 \\

$P$ (s)\dotfill         &   7.055217345(4)  \\
$\dot{P}$  (s s$^{-1}$)\dotfill   &   $3.0075(20)\times10^{-14}$  \\

$\dot{E}$ (erg s$^{-1}$)\dotfill & $3.38\times 10^{30}$ \\
$B_{\rm s}$ (G)\dotfill & $1.47\times10^{13}$ \\
$\tau_c$ (yr)\dotfill & $3.72\times10^{6}$ \\
         \bottomrule
\end{tabular}
\end{table}

\subsection{Energy-resolved pulse profiles}
\label{sec:pp}

\begin{figure}
    \centering
    \includegraphics[width=.45\textwidth]{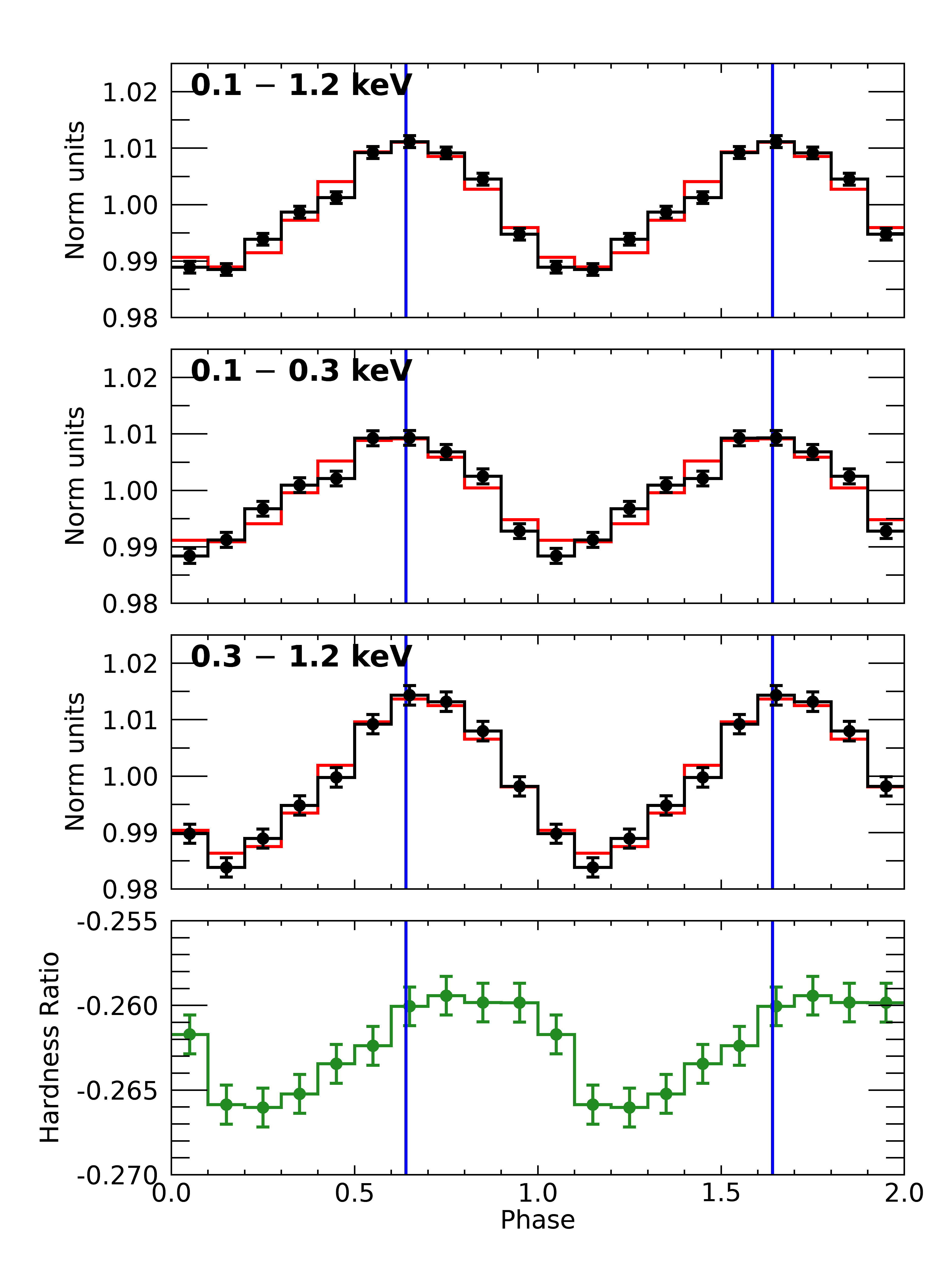}\\
    \includegraphics[trim=0cm 10cm 0cm 0.8cm,clip,width=.45\textwidth]{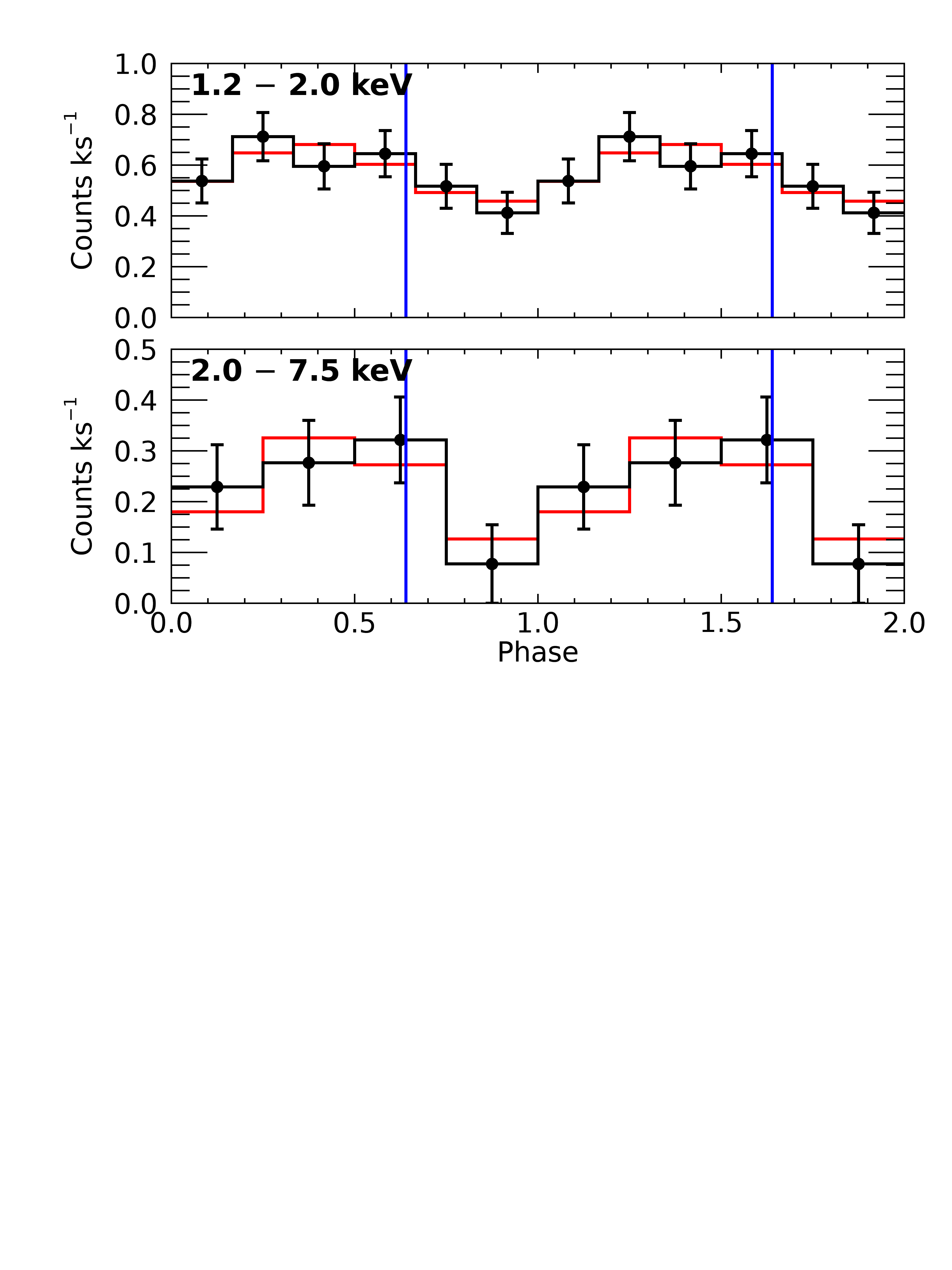}
    \caption{Pulse profiles between 0.1--1.2\,\si{\keV} (first panel), 0.1--{0.3}\,\si{\keV} (second panel), 0.3--1.2\,\si{\keV} (third panel), 1.2--2\,\si{\keV} (fifth panel), and 2--7.5\,\si{\keV} (sixth panel). The red lines show the best fit with a sinusoidal plus constant function, while the blue vertical lines show the phase corresponding to the peak ($\phi_0 = 0.640\pm0.007$). The fourth panel shows with green dots the hardness ratio between 0.1--0.3\,\si{\keV} and 0.3--1.2\,\si{\keV}  as a function of the phase.}
    \label{fig:lc_multi}
\end{figure}

We used the timing solution derived above to fold the counts of all the \xmm observations in order to obtain the pulse profile between 0.1--1.2\,\si{\keV}, shown in the upper panel of Fig.~\ref{fig:lc_multi}.
Its pulsed fraction is $\text{PF} \equiv (\text{CR}_{\rm max} - \text{CR}_{\rm min})/(\text{CR}_{\rm max} + \text{CR}_{\rm min}) = {(1.13\pm0.07)\%}$, where $\text{CR}$ is the background-subtracted count rate. A fit with a sinusoid plus a constant gives a $\chisq=21.9$ (for 7 dof, nhp = 0.0026, red line in Fig.~\ref{fig:lc_multi}).
The phase corresponding to the peak is $\phi_0=0.640 \pm 0.007$ (blue vertical line).

We repeated the analysis in the 0.1--0.3\,\si{\keV} and 0.3--1.2\,\si{\keV} energy ranges (second and third panel of Fig.~\ref{fig:lc_multi}). The softer profile has a smaller PF = $(1.04\pm0.09)$\% and is not consistent with a sinusoid ($\chi^2=20.6$ for 7 dof, nhp = 0.0044), as it is more skewed; its peak trails $\phi_0$ of $0.035\pm0.012$ cycles. Contrariwise, the harder profile, with  PF = $(1.53\pm0.12)$\%, is consistent with a sinusoid ($\chi^2=6.2$ for 7 dof, nhp = 0.52), and its peak leads $\phi_0$ of $0.041\pm0.011$ cycles.  
The different behaviour of the two   profiles can be seen in the fourth panel of  Fig.~\ref{fig:lc_multi}, where we plot the hardness ratio HR = (CR$_{0.3-1.2}$ -- CR$_{0.1-0.3}$)/(CR$_{0.3-1.2}$ + CR$_{0.1-0.3}$) as a function of the phase. The 0.1--0.3\,\si{\keV} range is rather narrow compared to the instrumental resolution $\Delta E\sim100$ eV (FWHM) at these energies, meaning that this hardness ratio does not reflect exactly the relative intensity of the flux in the two energy ranges. However, this systematic effect reduces the phase-modulation of the hardness ratio, compared to that obtained with a perfect instrument. We therefore conclude that there is  evidence for a slight phase-dependent spectral variation of the soft component.

We extracted the pulse profiles in the 1.2--2\,\si{\keV} and 2--7\,\si{\keV} using the ML method to derive the net source counts from images accumulated in 6 and 4 phase bins, respectively (see the two lower panels of Fig.~\ref{fig:lc_multi}). 
The energy range 1.2--2\,\si{\keV} has PF $=(29\pm11)\%$ and $\phi_0 = 0.39\pm0.16$.
Fitting these profiles with a constant yields a 
$\chisq=8.14$ for 5 dof and $\chisq=5.64$ for 3 dof, respectively. 
Although these values are not large enough to claim a significant detection of pulsations, the shape of the profiles and their broad alignment with those seen at lower energies suggest that also the flux at these energies is modulated at the spin period.

\section{Discussion} \label{sec:discussion}

Thanks to the combined use of \nicer and \xmm which provided, respectively, observations with frequent sampling and spanning a long time period, we could obtain a phase-coherent timing solution for \eighteen extending over 20 years. The only previous measurement of the spin-down rate of this XDINS had been reported by \citet{2008ApJ...673L.163V}. 
These authors used \xmm and \cha observations taken from 2000 to 2007 to set a 2$\sigma$ upper limit of |$\dot{\nu}$| $<1.3\times10^{-14}$ Hz s$^{-1}$, through a non-coherent timing analysis. The derivation of a univocal phase-coherent timing solution was hampered by the large separation between the observations, which led to some ambiguities in the count of the star rotation cycles. Nevertheless, \citet{2008ApJ...673L.163V} could derive the  most likely phase-coherent solution, yielding 
$\dot{\nu}$ = $(-5.98\pm0.14)\times10^{-16}$ Hz s$^{-1}$. 
The new data analysed here allowed us to derive, unambiguously and without loss of phase-coherence, a much more precise measurement of the spin-down rate,  $\dot{\nu}$ = $(-6.042\pm0.004)\times10^{-16}$ Hz s$^{-1}$. This value confirms the results by \citet{2008ApJ...673L.163V} and consolidates the estimates of rotational energy loss rate ($\dot{E}=\num{3.95e46}\dot{P}/P^3\,\si{\erg\per\second}=\SI{3.38e30}{\erg\per\second}$) and dipolar magnetic field ($B=\num{3.20e19}\sqrt{P/\dot{P}}\,\si{\gauss}=\SI{1.47e13}{\gauss}$).
Moreover, we find that the spin down has been remarkably stable over the two decades covered by our study (as proven by the very consistent spread of the residuals in Fig.\ \ref{sec:timing_sol} over time); this is in contrast to the other well studied XDINS {RX~J0720.4$-$23125}, which shows long-term spectral and timing variability with a timescale of a year \citep[and references therein]{2012MNRAS.423.1194H}.

Our $\dot{\nu}$ measurement is consistent with the picture of J1856 as a typical member of the XDINS class. The characteristic age $\tau_c=P/2\dot{P}=3.7\,\si{\mega\year}$ obtained from the dipole spin-down model is an order of magnitude higher than the kinematic age of $0.46 \pm 0.05\,\si{\mega\year}$, obtained by tracing the proper motion of J1856 back to its likely birthplace in the Upper Scorpius association \citep{2011MNRAS.417..617T,2013MNRAS.429.3517M}. 
This is a manifestation of the dissipation and evolution of the magnetic field \citep[e.g.][and references therein]{2021Univ....7..351I}, which is in line with the idea that XDINSs are older, worn-out magnetars.

Our spectral analysis confirms  that J1856 is detected also above $\approx$1\,keV  and its spectrum is more complex than a single soft blackbody-like component. 
In fact, a hard excess has been detected up to $\approx$8\,keV. It can be described with the sum of two components: a hot thermal one, dominating up to $\approx$2\,keV, and a power law emerging at higher energy (see the results in Table~\ref{tab:hard_components}). These two components had been separately detected in previous works: \citet{2017PASJ...69...50Y} found that the excess around $\approx$1\,keV could be accounted for by a blackbody with $kT^\infty=137^{+18}_{-14}\,\si{\eV}$, $R^\infty=36^{+45}_{-36}\,\si{\meter}$ whereas  \citet{2020ApJ...904...42D} found a power law in the $2$ and \SI{8}{\keV} band with {$\Gamma=1.0^{+1.1}_{-0.9}$}, $I_{\text{[2--8]}}=(2.1\pm 0.9)\times10^{-15}$\,\si{\flux}. We found that indeed both components are needed at the same time, and the parameter of our best fit are in good agreement with those reported by these authors.

The luminosity of the non-thermal component corresponds to $10^{-3}$ times the spin-down power $\dot{E}$. This value is consistent with what is observed in rotation powered X-ray pulsars with higher $\dot{E}$ \citep[e.g.][]{2002A&A...387..993P}, so that a magnetospheric origin for this component appears as the most natural option.
Nevertheless, the lack of a radio counterpart
has recently led 
to more exotic suggestions, like the production of axion-like particles in the stellar core, to explain the hard power-law tail in J1856 \citep{2021PhRvL.126b1102B}. This may be tested for all the seven XDINS, and indeed \citet{2020ApJ...904...42D} found evidence of a power-law tail in the spectrum of {RX~J0420.0$-$5022}. We report in Appendix \ref{sec:J0420} our spectral analysis of this object. We find a hard excess with respect to a single component fit, which may be interpreted as a second blackbody in accordance with the results by \citet{2019PASJ...71...17Y}, who found evidence of harder thermal components in all the XDINSs. However, the amount of data is not sufficient to constrain the nature of this hard excess, and in particular the presence of a non-thermal component at even higher energy cannot be unambiguously established. We nevertheless note that {RX~J0420.0$-$5022} and J1856 have the highest $\dot E$ of the class, so that they stand out as the most probable candidates to emit non thermally. Further data will be needed to assess whether non-thermal emission is common to all the members of the XDINS class, 
or J1856 (and possibly {RX~J0420.0$-$5022}) should be regarded as a separate NS flavour.

Of course, a significant detection of pulsations above $\sim$1.2\,\si{\keV} would give important clues on the origin of the hard X-rays.
If the pulsations between 1.2--2\,\si{\keV} with a PF of about 30\% were confirmed, 
it would be natural to interpret the harder blackbody component as thermal emission from a hot spot on the star surface. 
The current data do not allow to draw firm conclusions, although the pulse profiles shown in Fig.~\ref{fig:lc_multi} are intriguing. 
On the other hand, thanks to the large exposure time collected with \xmm we discovered a statistically significant variation in the 0.1--0.3\,\si{\keV} to 0.3--1.2\,\si{\keV} hardness ratio, which indicates that the spectrum gets harder at the peak of the pulse profile.
Furthermore, we found that the 0.1--0.3\,\si{\keV} pulse profile is somewhat skewed.
These effects are noticeable below $\approx$1\,keV, where the cooler blackbody dominates, and imply a certain degree of anisotropy in the thermal surface distribution and/or in the angular emission pattern. The dependence of the phase of the pulse maximum on the energy we found in J1856 bears some resemblance to what was reported in the magnetar XTE J1810$-$197 in outburst \citep{2021MNRAS.504.5244B} and may be again indicative of a non-axially symmetric thermal map. 

Throughout this work, we described the thermal components of J1856 using blackbody models. This was not done merely for the sake of simplicity: we also tested models of emission from a magnetised atmosphere (namely, the \textsc{NSMAXG} models \citealp{2008ApJS..178..102H} for $B=\SI{e13}{\gauss}$), finding that they consistently performed more poorly than the blackbody ($\chisq/\text{~dof}\gg 10$ in all instances). In fact, atmosphere models tend to produce a broader thermal spectrum throughout, rather than increasing the hard emission only. This points to the surface of J1856 being in some kind of condensed state, which shows an emission that is very close to a blackbody \citep{2005ApJ...628..902V}. Moreover, \citet{2022MNRAS.513.3113R} showed that a relatively steep thermal map and atmospheric hardening produce a thermal spectrum that can be well fitted with two blackbody components, as long as $T_2/T_1\sim2$ and $R_2/R_1\sim0.1$. These values are suitable for the majority of the thermally-emitting isolated NSs, with the exception of PSR\,J0633+1746 (Geminga), RX\,J0420.0$-$5022 and J1856 itself.

We finally note that the emission detected from J1856 in the optical band, consistent in with a Rayleigh tail \citep{2001A&A...378..986V, 2002ApJ...564..981P}, requires the presence of a third, colder blackbody component. Indeed, our best fit model extrapolated to the optical/UV band still predicts a flux much lower than the observed one. At the same time, as found also in \sartore, the inclusion of an additional blackbody component with $kT^\infty\simeq\SI{30}{\eV}$ and emission area corresponding to the majority of the stellar surface ($R^\infty\simeq\SI{16}{\kilo\meter}$) accounting for the optical flux, alters only marginally our best fit parameters of the X-ray components.

\section{Conclusions} \label{sec:conclusions}

Our analysis of 1.43 Ms of data obtained with \xmm over the last 20 years showed that the X-ray spectrum of J1856 is more complex than originally believed and confirms independently reported evidence for emission above $\sim$1.2\,\si{\keV}.  When coupled to the optical/UV observations, these data point to a neutron star shining mainly by thermal emission from a large fraction of its surface at an observed temperature of $\sim$62 eV. The remaining part of the star surface is a factor $\sim$2 cooler and is visible in the optical/UV band. We note that the ratios of these temperatures ($\sim$2) and emitting radii ($\sim$0.1) fit well with those of other thermally-emitting NSs as shown in Fig. 7 of \citet{2022MNRAS.513.3113R}.
The picture is complicated by the presence of a further hotter thermal component and a power-law tail. The former could result from a relatively small hot spot, consistent with the marginal evidence for pulsations in the 1.2--2\,\si{\keV} range, while the latter, if interpreted as non-thermal emission of magnetospheric origin, implies an efficiency in line with that observed for pulsars with higher $\dot E$.
As the NS is isolated, the three thermal components are most likely an approximation of a more complex continuous thermal distribution on the surface resulting from its magnetothermal evolution; in particular, this points to a multipolar field structure \citep[see e.g.\ the model proposed in][]{2021ApJ...914..118D}.

A phase-connected timing analysis shows a remarkable stability of the spin-down rate of J1856. Also its flux and spectrum remained virtually unchanged over a 20 years time span. Such properties were thought  until recently to be shared by all the other 
XDINSs, with the notable exception of {RX~J0720.4$-$3125} \citep{2012MNRAS.423.1194H}.
However, new observations obtained with \textit{eROSITA} \citep{2022arXiv220206793M} indicate the presence of long term variability in the spectral and timing properties of two more members of this class.
These findings, together with the evidence that
J1856 (and possibly {RX~J0420.0$-$5022}) are not purely-thermally emitting NS, indicate that,
notwithstanding the small number of its representatives, the category of XDINSs is not so homogeneous. Future observations will have to appraise the merits and shortcomings of the XDINSs classification.


\section*{Acknowledgements}
The  results reported in this article are based on data obtained with \xmm, an ESA science mission with instruments and contributions directly funded by ESA Member States and the USA (NASA) and by the NASA NICER mission and the Astrophysics Explorers Program. The data analysis has benefited from data and software provided by the High Energy Astrophysics Science Archive Research Center (HEASARC), a service of the Astrophysics Science Division at NASA/GSFC and the High Energy Astrophysics Division of the Smithsonian Astrophysical Observatory. 
We acknowledge financial support from the Italian Ministry for University and Research through grant UnIAM (2017LJ39LM, PI S.~Mereghetti). G.~Y. research is supported by an appointment to the NASA Postdoctoral Program at the Goddard Space Flight Center, administered by Oak Ridge Associated Universities under contract with NASA.

\section*{Data availability}
All the data used in this article are available in public archives.

\bibliographystyle{mnras}
\bibliography{biblio}

\begin{thebibliography}{}
\makeatletter
\relax
\def\mn@urlcharsother{\let\do\@makeother \do\$\do\&\do\#\do\^\do\_\do\%\do\~}
\def\mn@doi{\begingroup\mn@urlcharsother \@ifnextchar [ {\mn@doi@}
  {\mn@doi@[]}}
\def\mn@doi@[#1]#2{\def\@tempa{#1}\ifx\@tempa\@empty \href
  {http://dx.doi.org/#2} {doi:#2}\else \href {http://dx.doi.org/#2} {#1}\fi
  \endgroup}
\def\mn@eprint#1#2{\mn@eprint@#1:#2::\@nil}
\def\mn@eprint@arXiv#1{\href {http://arxiv.org/abs/#1} {{\tt arXiv:#1}}}
\def\mn@eprint@dblp#1{\href {http://dblp.uni-trier.de/rec/bibtex/#1.xml}
  {dblp:#1}}
\def\mn@eprint@#1:#2:#3:#4\@nil{\def\@tempa {#1}\def\@tempb {#2}\def\@tempc
  {#3}\ifx \@tempc \@empty \let \@tempc \@tempb \let \@tempb \@tempa \fi \ifx
  \@tempb \@empty \def\@tempb {arXiv}\fi \@ifundefined
  {mn@eprint@\@tempb}{\@tempb:\@tempc}{\expandafter \expandafter \csname
  mn@eprint@\@tempb\endcsname \expandafter{\@tempc}}}

\bibitem[\protect\citeauthoryear{{\VAN{Adelsberg}{van}{van} Adelsberg}, {Lai},
  {Potekhin}  \& {Arras}}{{\VAN{Adelsberg}{van}{van} Adelsberg}
  et~al.}{2005}]{2005ApJ...628..902V}
{\VAN{Adelsberg}{van}{van} Adelsberg} M.,  {Lai} D.,  {Potekhin} A.~Y.,
  {Arras} P.,  2005, \mn@doi [\apj] {10.1086/430871}, \href
  {https://ui.adsabs.harvard.edu/abs/2005ApJ...628..902V} {628, 902}

\bibitem[\protect\citeauthoryear{{Beuermann}, {Burwitz}  \&
  {Rauch}}{{Beuermann} et~al.}{2006}]{2006A&A...458..541B}
{Beuermann} K.,  {Burwitz} V.,   {Rauch} T.,  2006, \mn@doi [\aap]
  {10.1051/0004-6361:20065478}, \href
  {https://ui.adsabs.harvard.edu/abs/2006A&A...458..541B} {458, 541}

\bibitem[\protect\citeauthoryear{{Borghese}, {Rea}, {Coti Zelati}, {Tiengo}  \&
  {Turolla}}{{Borghese} et~al.}{2015}]{2015ApJ...807L..20B}
{Borghese} A.,  {Rea} N.,  {Coti Zelati} F.,  {Tiengo} A.,   {Turolla} R.,
  2015, \mn@doi [\apjl] {10.1088/2041-8205/807/1/L20}, \href
  {https://ui.adsabs.harvard.edu/abs/2015ApJ...807L..20B} {807, L20}

\bibitem[\protect\citeauthoryear{{Borghese}, {Rea}, {Coti Zelati}, {Tiengo},
  {Turolla}  \& {Zane}}{{Borghese} et~al.}{2017}]{2017MNRAS.468.2975B}
{Borghese} A.,  {Rea} N.,  {Coti Zelati} F.,  {Tiengo} A.,  {Turolla} R.,
  {Zane} S.,  2017, \mn@doi [\mnras] {10.1093/mnras/stx632}, \href
  {https://ui.adsabs.harvard.edu/abs/2017MNRAS.468.2975B} {468, 2975}

\bibitem[\protect\citeauthoryear{{Borghese} et~al.,}{{Borghese}
  et~al.}{2021}]{2021MNRAS.504.5244B}
{Borghese} A.,  et~al., 2021, \mn@doi [\mnras] {10.1093/mnras/stab1236}, \href
  {https://ui.adsabs.harvard.edu/abs/2021MNRAS.504.5244B} {504, 5244}

\bibitem[\protect\citeauthoryear{{Buccheri} et~al.,}{{Buccheri}
  et~al.}{1983}]{1983A&A...128..245B}
{Buccheri} R.,  et~al., 1983, \aap, \href
  {https://ui.adsabs.harvard.edu/abs/1983A&A...128..245B} {128, 245}

\bibitem[\protect\citeauthoryear{{Burwitz}, {Zavlin}, {Neuh{\"a}user},
  {Predehl}, {Tr{\"u}mper}  \& {Brinkman}}{{Burwitz}
  et~al.}{2001}]{2001A&A...379L..35B}
{Burwitz} V.,  {Zavlin} V.~E.,  {Neuh{\"a}user} R.,  {Predehl} P.,
  {Tr{\"u}mper} J.,   {Brinkman} A.~C.,  2001, \mn@doi [\aap]
  {10.1051/0004-6361:20011304}, \href
  {https://ui.adsabs.harvard.edu/abs/2001A&A...379L..35B} {379, L35}

\bibitem[\protect\citeauthoryear{{Burwitz}, {Haberl}, {Neuh{\"a}user},
  {Predehl}, {Tr{\"u}mper}  \& {Zavlin}}{{Burwitz}
  et~al.}{2003}]{2003A&A...399.1109B}
{Burwitz} V.,  {Haberl} F.,  {Neuh{\"a}user} R.,  {Predehl} P.,  {Tr{\"u}mper}
  J.,   {Zavlin} V.~E.,  2003, \mn@doi [\aap] {10.1051/0004-6361:20021747},
  \href {https://ui.adsabs.harvard.edu/abs/2003A&A...399.1109B} {399, 1109}

\bibitem[\protect\citeauthoryear{{Buschmann}, {Co}, {Dessert}  \&
  {Safdi}}{{Buschmann} et~al.}{2021}]{2021PhRvL.126b1102B}
{Buschmann} M.,  {Co} R.~T.,  {Dessert} C.,   {Safdi} B.~R.,  2021, \mn@doi
  [\prl] {10.1103/PhysRevLett.126.021102}, \href
  {https://ui.adsabs.harvard.edu/abs/2021PhRvL.126b1102B} {126, 021102}

\bibitem[\protect\citeauthoryear{{Dall'Osso}, {Israel}, {Stella}, {Possenti}
  \& {Perozzi}}{{Dall'Osso} et~al.}{2003}]{2003ApJ...599..485D}
{Dall'Osso} S.,  {Israel} G.~L.,  {Stella} L.,  {Possenti} A.,   {Perozzi} E.,
  2003, \mn@doi [\apj] {10.1086/379213}, \href
  {https://ui.adsabs.harvard.edu/abs/2003ApJ...599..485D} {599, 485}

\bibitem[\protect\citeauthoryear{{De Grandis}, {Taverna}, {Turolla}, {Gnarini},
  {Popov}, {Zane}  \& {Wood}}{{De Grandis} et~al.}{2021}]{2021ApJ...914..118D}
{De Grandis} D.,  {Taverna} R.,  {Turolla} R.,  {Gnarini} A.,  {Popov} S.~B.,
  {Zane} S.,   {Wood} T.~S.,  2021, \mn@doi [\apj] {10.3847/1538-4357/abfdac},
  \href {https://ui.adsabs.harvard.edu/abs/2021ApJ...914..118D} {914, 118}

\bibitem[\protect\citeauthoryear{{Dessert}, {Foster}  \& {Safdi}}{{Dessert}
  et~al.}{2020}]{2020ApJ...904...42D}
{Dessert} C.,  {Foster} J.~W.,   {Safdi} B.~R.,  2020, \mn@doi [\apj]
  {10.3847/1538-4357/abb4ea}, \href
  {https://ui.adsabs.harvard.edu/abs/2020ApJ...904...42D} {904, 42}

\bibitem[\protect\citeauthoryear{{Ghizzardi}}{{Ghizzardi}}{2002}]{simo}
{Ghizzardi} S.,  2002, {In flight calibration of the PSF for the PN camera}.
{XMM-SOC-CAL-TN-0029},
  http://www.cosmos.esa.int/web/xmm-newton/calibration-documentation

\bibitem[\protect\citeauthoryear{{Haberl}, {Schwope}, {Hambaryan}, {Hasinger}
  \& {Motch}}{{Haberl} et~al.}{2003}]{2003A&A...403L..19H}
{Haberl} F.,  {Schwope} A.~D.,  {Hambaryan} V.,  {Hasinger} G.,   {Motch} C.,
  2003, \mn@doi [\aap] {10.1051/0004-6361:20030450}, \href
  {https://ui.adsabs.harvard.edu/abs/2003A&A...403L..19H} {403, L19}

\bibitem[\protect\citeauthoryear{{Haberl} et~al.,}{{Haberl}
  et~al.}{2004}]{2004A&A...424..635H}
{Haberl} F.,  et~al., 2004, \mn@doi [\aap] {10.1051/0004-6361:20040440}, \href
  {https://ui.adsabs.harvard.edu/abs/2004A&A...424..635H} {424, 635}

\bibitem[\protect\citeauthoryear{{Ho}, {Kaplan}, {Chang}, {van Adelsberg}  \&
  {Potekhin}}{{Ho} et~al.}{2007}]{2007MNRAS.375..821H}
{Ho} W. C.~G.,  {Kaplan} D.~L.,  {Chang} P.,  {van Adelsberg} M.,   {Potekhin}
  A.~Y.,  2007, \mn@doi [\mnras] {10.1111/j.1365-2966.2006.11376.x}, \href
  {https://ui.adsabs.harvard.edu/abs/2007MNRAS.375..821H} {375, 821}

\bibitem[\protect\citeauthoryear{{Ho}, {Potekhin}  \& {Chabrier}}{{Ho}
  et~al.}{2008}]{2008ApJS..178..102H}
{Ho} W. C.~G.,  {Potekhin} A.~Y.,   {Chabrier} G.,  2008, \mn@doi [\apjs]
  {10.1086/589238}, \href
  {https://ui.adsabs.harvard.edu/abs/2008ApJS..178..102H} {178, 102}

\bibitem[\protect\citeauthoryear{{Hohle}, {Haberl}, {Vink}, {de Vries},
  {Turolla}, {Zane}  \& {M{\'e}ndez}}{{Hohle}
  et~al.}{2012}]{2012MNRAS.423.1194H}
{Hohle} M.~M.,  {Haberl} F.,  {Vink} J.,  {de Vries} C.~P.,  {Turolla} R.,
  {Zane} S.,   {M{\'e}ndez} M.,  2012, \mn@doi [\mnras]
  {10.1111/j.1365-2966.2012.20946.x}, \href
  {https://ui.adsabs.harvard.edu/abs/2012MNRAS.423.1194H} {423, 1194}

\bibitem[\protect\citeauthoryear{{Igoshev}, {Popov}  \& {Hollerbach}}{{Igoshev}
  et~al.}{2021}]{2021Univ....7..351I}
{Igoshev} A.~P.,  {Popov} S.~B.,   {Hollerbach} R.,  2021, \mn@doi [Universe]
  {10.3390/universe7090351}, \href
  {https://ui.adsabs.harvard.edu/abs/2021Univ....7..351I} {7, 351}

\bibitem[\protect\citeauthoryear{{\VAN{Kerkwijk}{van}{van} Kerkwijk} \&
  {Kaplan}}{{\VAN{Kerkwijk}{van}{van} Kerkwijk} \&
  {Kaplan}}{2007}]{2007Ap&SS.308..191V}
{\VAN{Kerkwijk}{van}{van} Kerkwijk} M.~H.,  {Kaplan} D.~L.,  2007, \mn@doi
  [\apss] {10.1007/s10509-007-9343-9}, \href
  {https://ui.adsabs.harvard.edu/abs/2007Ap&SS.308..191V} {308, 191}

\bibitem[\protect\citeauthoryear{{\VAN{Kerkwijk}{van}{van} Kerkwijk} \&
  {Kaplan, D.~L.}}{{\VAN{Kerkwijk}{van}{van} Kerkwijk} \& {Kaplan,
  D.~L.}}{2008}]{2008ApJ...673L.163V}
{\VAN{Kerkwijk}{van}{van} Kerkwijk} M.~H.,  {Kaplan, D.~L.} 2008, \mn@doi
  [\apjl] {10.1086/528796}, \href
  {https://ui.adsabs.harvard.edu/abs/2008ApJ...673L.163V} {673, L163}

\bibitem[\protect\citeauthoryear{{\VAN{Kerkwijk}{van}{van} Kerkwijk} \&
  {Kulkarni}}{{\VAN{Kerkwijk}{van}{van} Kerkwijk} \&
  {Kulkarni}}{2001}]{2001A&A...378..986V}
{\VAN{Kerkwijk}{van}{van} Kerkwijk} M.~H.,  {Kulkarni} S.~R.,  2001, \mn@doi
  [\aap] {10.1051/0004-6361:20011272}, \href
  {https://ui.adsabs.harvard.edu/abs/2001A&A...378..986V} {378, 986}

\bibitem[\protect\citeauthoryear{{\VAN{Kerkwijk}{van}{van} Kerkwijk}, {Kaplan},
  {Durant}, {Kulkarni}  \& {Paerels}}{{\VAN{Kerkwijk}{van}{van} Kerkwijk}
  et~al.}{2004}]{2004ApJ...608..432V}
{\VAN{Kerkwijk}{van}{van} Kerkwijk} M.~H.,  {Kaplan} D.~L.,  {Durant} M.,
  {Kulkarni} S.~R.,   {Paerels} F.,  2004, \mn@doi [\apj] {10.1086/386299},
  \href {https://ui.adsabs.harvard.edu/abs/2004ApJ...608..432V} {608, 432}

\bibitem[\protect\citeauthoryear{{Livingstone}, {Ransom}, {Camilo}, {Kaspi},
  {Lyne}, {Kramer}  \& {Stairs}}{{Livingstone}
  et~al.}{2009}]{2009ApJ...706.1163L}
{Livingstone} M.~A.,  {Ransom} S.~M.,  {Camilo} F.,  {Kaspi} V.~M.,  {Lyne}
  A.~G.,  {Kramer} M.,   {Stairs} I.~H.,  2009, \mn@doi [\apj]
  {10.1088/0004-637X/706/2/1163}, \href
  {https://ui.adsabs.harvard.edu/abs/2009ApJ...706.1163L} {706, 1163}

\bibitem[\protect\citeauthoryear{{Mancini Pires}, {Schwope}  \&
  {Kurpas}}{{Mancini Pires} et~al.}{2022}]{2022arXiv220206793M}
{Mancini Pires} A.,  {Schwope} A.,   {Kurpas} J.,  2022, arXiv e-prints, \href
  {https://ui.adsabs.harvard.edu/abs/2022arXiv220206793M} {p. arXiv:2202.06793}

\bibitem[\protect\citeauthoryear{{Mignani} et~al.,}{{Mignani}
  et~al.}{2013}]{2013MNRAS.429.3517M}
{Mignani} R.~P.,  et~al., 2013, \mn@doi [\mnras] {10.1093/mnras/sts627}, \href
  {https://ui.adsabs.harvard.edu/abs/2013MNRAS.429.3517M} {429, 3517}

\bibitem[\protect\citeauthoryear{{Neuhaeuser}, {Thomas}, {Danner}, {Peschke}
  \& {Walter}}{{Neuhaeuser} et~al.}{1997}]{1997A&A...318L..43N}
{Neuhaeuser} R.,  {Thomas} H.~C.,  {Danner} R.,  {Peschke} S.,   {Walter}
  F.~M.,  1997, \aap, \href
  {https://ui.adsabs.harvard.edu/abs/1997A&A...318L..43N} {318, L43}

\bibitem[\protect\citeauthoryear{{Pavlov}, {Zavlin}, {Truemper}  \&
  {Neuhaeuser}}{{Pavlov} et~al.}{1996}]{1996ApJ...472L..33P}
{Pavlov} G.~G.,  {Zavlin} V.~E.,  {Truemper} J.,   {Neuhaeuser} R.,  1996,
  \mn@doi [\apjl] {10.1086/310355}, \href
  {https://ui.adsabs.harvard.edu/abs/1996ApJ...472L..33P} {472, L33}

\bibitem[\protect\citeauthoryear{{Pons}, {Walter}, {Lattimer}, {Prakash},
  {Neuh{\"a}user}  \& {An}}{{Pons} et~al.}{2002}]{2002ApJ...564..981P}
{Pons} J.~A.,  {Walter} F.~M.,  {Lattimer} J.~M.,  {Prakash} M.,
  {Neuh{\"a}user} R.,   {An} P.,  2002, \mn@doi [\apj] {10.1086/324296}, \href
  {https://ui.adsabs.harvard.edu/abs/2002ApJ...564..981P} {564, 981}

\bibitem[\protect\citeauthoryear{{Posselt}, {Popov}, {Haberl}, {Tr{\"u}mper},
  {Turolla}  \& {Neuh{\"a}user}}{{Posselt} et~al.}{2007}]{2007Ap&SS.308..171P}
{Posselt} B.,  {Popov} S.~B.,  {Haberl} F.,  {Tr{\"u}mper} J.,  {Turolla} R.,
  {Neuh{\"a}user} R.,  2007, \mn@doi [\apss] {10.1007/s10509-007-9344-8}, \href
  {https://ui.adsabs.harvard.edu/abs/2007Ap&SS.308..171P} {308, 171}

\bibitem[\protect\citeauthoryear{{Possenti}, {Cerutti}, {Colpi}  \&
  {Mereghetti}}{{Possenti} et~al.}{2002}]{2002A&A...387..993P}
{Possenti} A.,  {Cerutti} R.,  {Colpi} M.,   {Mereghetti} S.,  2002, \mn@doi
  [\aap] {10.1051/0004-6361:20020472}, \href
  {https://ui.adsabs.harvard.edu/abs/2002A&A...387..993P} {387, 993}

\bibitem[\protect\citeauthoryear{{Ray} et~al.,}{{Ray}
  et~al.}{2011}]{2011ApJS..194...17R}
{Ray} P.~S.,  et~al., 2011, \mn@doi [\apjs] {10.1088/0067-0049/194/2/17}, \href
  {https://ui.adsabs.harvard.edu/abs/2011ApJS..194...17R} {194, 17}

\bibitem[\protect\citeauthoryear{{Read}, {Guainazzi}  \& {Sembay}}{{Read}
  et~al.}{2014}]{2014A&A...564A..75R}
{Read} A.~M.,  {Guainazzi} M.,   {Sembay} S.,  2014, \mn@doi [\aap]
  {10.1051/0004-6361/201423422}, \href
  {https://ui.adsabs.harvard.edu/abs/2014A&A...564A..75R} {564, A75}

\bibitem[\protect\citeauthoryear{{Rigoselli}, {Mereghetti}, {Taverna},
  {Turolla}  \& {De Grandis}}{{Rigoselli} et~al.}{2021}]{2021A&A...646A.117R}
{Rigoselli} M.,  {Mereghetti} S.,  {Taverna} R.,  {Turolla} R.,   {De Grandis}
  D.,  2021, \mn@doi [\aap] {10.1051/0004-6361/202039774}, \href
  {https://ui.adsabs.harvard.edu/abs/2021A&A...646A.117R} {646, A117}

\bibitem[\protect\citeauthoryear{{Rigoselli}, {Mereghetti}, {Anzuinelli},
  {Keith}, {Taverna}, {Turolla}  \& {Zane}}{{Rigoselli}
  et~al.}{2022}]{2022MNRAS.513.3113R}
{Rigoselli} M.,  {Mereghetti} S.,  {Anzuinelli} S.,  {Keith} M.,  {Taverna} R.,
   {Turolla} R.,   {Zane} S.,  2022, \mn@doi [\mnras] {10.1093/mnras/stac1130},
  \href {https://ui.adsabs.harvard.edu/abs/2022MNRAS.513.3113R} {513, 3113}

\bibitem[\protect\citeauthoryear{{Sartore}, {Tiengo}, {Mereghetti}, {De Luca},
  {Turolla}  \& {Haberl}}{{Sartore} et~al.}{2012}]{2012AA...541A..66S}
{Sartore} N.,  {Tiengo} A.,  {Mereghetti} S.,  {De Luca} A.,  {Turolla} R.,
  {Haberl} F.,  2012, \mn@doi [\aap] {10.1051/0004-6361/201118489}, \href
  {https://ui.adsabs.harvard.edu/abs/2012A&A...541A..66S} {541, A66}

\bibitem[\protect\citeauthoryear{{Tetzlaff}, {Eisenbeiss}, {Neuh{\"a}user}  \&
  {Hohle}}{{Tetzlaff} et~al.}{2011}]{2011MNRAS.417..617T}
{Tetzlaff} N.,  {Eisenbeiss} T.,  {Neuh{\"a}user} R.,   {Hohle} M.~M.,  2011,
  \mn@doi [\mnras] {10.1111/j.1365-2966.2011.19302.x}, \href
  {https://ui.adsabs.harvard.edu/abs/2011MNRAS.417..617T} {417, 617}

\bibitem[\protect\citeauthoryear{{Tiengo} \& {Mereghetti}}{{Tiengo} \&
  {Mereghetti}}{2007}]{2007ApJ...657L.101T}
{Tiengo} A.,  {Mereghetti} S.,  2007, \mn@doi [\apjl] {10.1086/513143}, \href
  {https://ui.adsabs.harvard.edu/abs/2007ApJ...657L.101T} {657, L101}

\bibitem[\protect\citeauthoryear{{Turolla}}{{Turolla}}{2009}]{2009ASSL..357..141T}
{Turolla} R.,  2009, {Isolated Neutron Stars: The Challenge of Simplicity}.
{Becker}, Werner, p.~141, \mn@doi{10.1007/978-3-540-76965-1_7}

\bibitem[\protect\citeauthoryear{{Walter}, {Wolk}  \& {Neuh{\"a}user}}{{Walter}
  et~al.}{1996}]{1996Natur.379..233W}
{Walter} F.~M.,  {Wolk} S.~J.,   {Neuh{\"a}user} R.,  1996, \mn@doi [\nat]
  {10.1038/379233a0}, \href
  {https://ui.adsabs.harvard.edu/abs/1996Natur.379..233W} {379, 233}

\bibitem[\protect\citeauthoryear{{Walter}, {Eisenbei{\ss}}, {Lattimer}, {Kim},
  {Hambaryan}  \& {Neuh{\"a}user}}{{Walter} et~al.}{2010}]{2010ApJ...724..669W}
{Walter} F.~M.,  {Eisenbei{\ss}} T.,  {Lattimer} J.~M.,  {Kim} B.,  {Hambaryan}
  V.,   {Neuh{\"a}user} R.,  2010, \mn@doi [\apj]
  {10.1088/0004-637X/724/1/669}, \href
  {https://ui.adsabs.harvard.edu/abs/2010ApJ...724..669W} {724, 669}

\bibitem[\protect\citeauthoryear{{Yoneyama}, {Hayashida}, {Nakajima}, {Inoue}
  \& {Tsunemi}}{{Yoneyama} et~al.}{2017}]{2017PASJ...69...50Y}
{Yoneyama} T.,  {Hayashida} K.,  {Nakajima} H.,  {Inoue} S.,   {Tsunemi} H.,
  2017, \mn@doi [\pasj] {10.1093/pasj/psx025}, \href
  {https://ui.adsabs.harvard.edu/abs/2017PASJ...69...50Y} {69, 50}

\bibitem[\protect\citeauthoryear{{Yoneyama}, {Hayashida}, {Nakajima}  \&
  {Matsumoto}}{{Yoneyama} et~al.}{2019}]{2019PASJ...71...17Y}
{Yoneyama} T.,  {Hayashida} K.,  {Nakajima} H.,   {Matsumoto} H.,  2019,
  \mn@doi [\pasj] {10.1093/pasj/psy135}, \href
  {https://ui.adsabs.harvard.edu/abs/2019PASJ...71...17Y} {71, 17}

\bibitem[\protect\citeauthoryear{{Zane}, {Cropper}, {Turolla}, {Zampieri},
  {Chieregato}, {Drake}  \& {Treves}}{{Zane}
  et~al.}{2005}]{2005ApJ...627..397Z}
{Zane} S.,  {Cropper} M.,  {Turolla} R.,  {Zampieri} L.,  {Chieregato} M.,
  {Drake} J.~J.,   {Treves} A.,  2005, \mn@doi [\apj] {10.1086/430138}, \href
  {https://ui.adsabs.harvard.edu/abs/2005ApJ...627..397Z} {627, 397}

\makeatother
\end{thebibliography}

\appendix

\section{Spectral analysis of the XDINS RX\,J0420.0$-$5022}
\label{sec:J0420}

We analysed 19 EPIC-pn observations of RX\,J0420.0$-$5022, which span from 2002 to 2019 (see Table~\ref{tab:observations_J04}). We followed the same data reduction processes described in Section~\ref{sec:data} and obtained 144 ks of net exposure time.

The source is well detected up to $\approx$0.5\,\si{\keV}, while at higher energies the spectrum is background-dominated. After checking that the bulk of the emission is constant throughout time, we stacked all the observations to produce summed images in the 0.5--1\,\si{\keV} and 1--5\,\si{\keV} ranges. 
Applying the ML source detection to these images in the two bands, we obtained $840\pm38$ (38.5$\sigma$) and $173\pm25$ (8.7$\sigma$) net source counts, respectively.

\begin{figure}
    \centering
    \includegraphics[width=.48\textwidth]{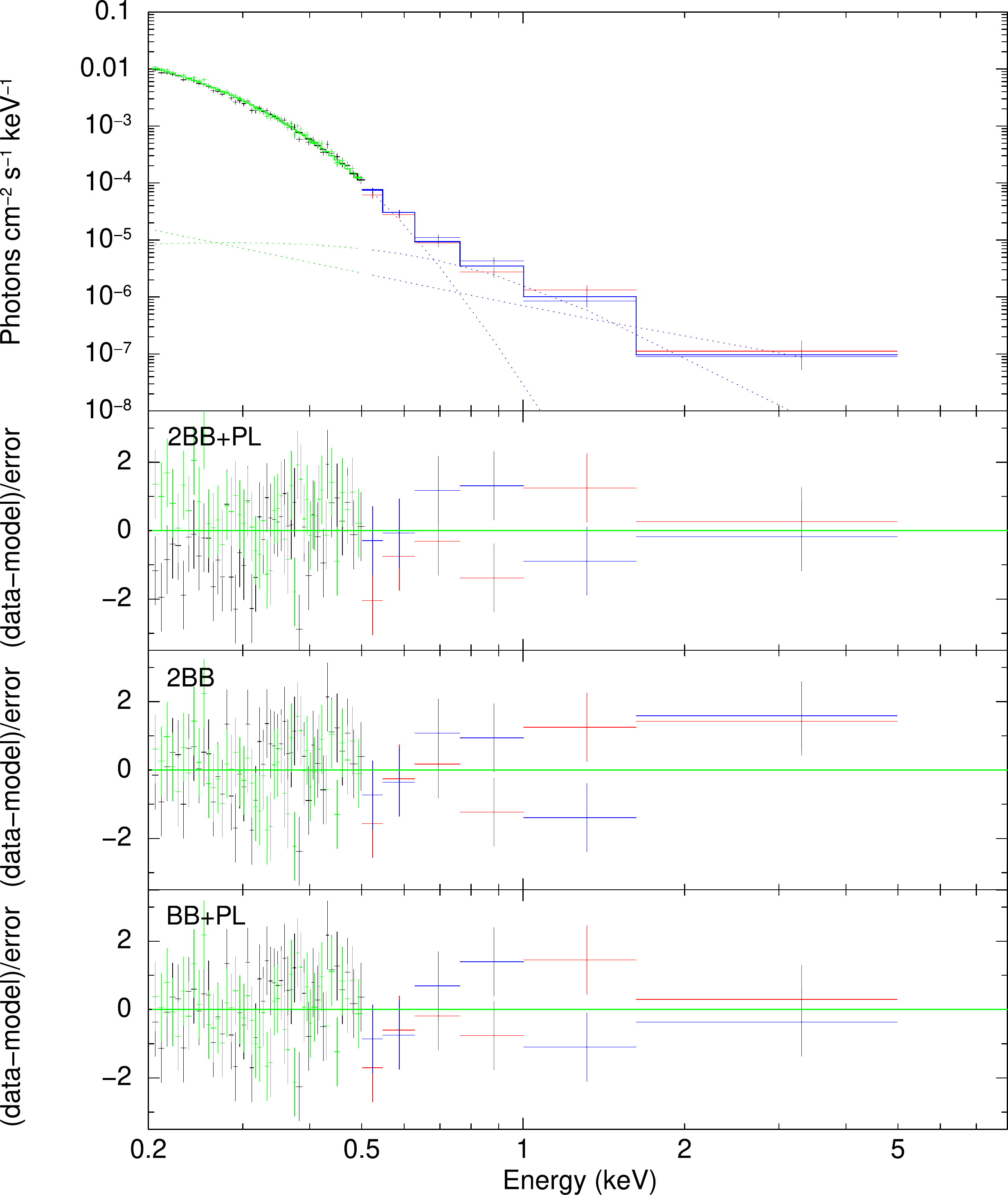}
    \caption{Same as Fig.\ \ref{fig:spec_all} but for {RX~J0420.0$-$5022}. The black/red points and curves correspond to the observations done in Small window mode, whereas the green/blue ones to the observations in Full frame.}
    \label{fig:0420}
\end{figure}

Then, we extracted the spectra taking care to group the observations according to the science mode in which the EPIC-pn was operating, in order to avoid cross-calibration uncertainties. The 5 observations taken in small window and the 14 observations taken in full frame provide  similar exposure times: 68.5 ks and 75.8 ks, respectively. As we did for J1856, we extracted the soft spectrum (below 0.5\,\si{\keV}) with a standard analysis selecting only events with $\textsc{pattern} = 0$, whereas we extracted the hard spectrum (0.5--5\,\si{\keV}) with the ML technique (and $\textsc{pattern} \leq 4$). The resulting spectrum is shown in Fig.\ \ref{fig:0420}.

We fitted simultaneously the four spectra (two for each science mode, divided over two energy ranges), finding that, in addition to the soft component fitted by a blackbody ($N_{\rm H}<10^{19}$ cm$^{-2}$, $kT_1^\infty=46.3_{-0.5}^{+0.3}$ eV, $R_1^\infty=4.6_{-0.1}^{+0.2}$ km\footnote{We used a distance of 335 pc, that is within the range 325--345 pc given by \citet{2007Ap&SS.308..171P}.}) there is a hard excess that can be fitted either with a second blackbody ($kT_2^\infty=210_{-20}^{+30}$ eV, $R_2^\infty=17_{-4}^{+3}$ m, $\chisq = 94.36$ with 100 dof) or with a power law ($\Gamma=3.10\pm0.25$, $I_{\text{[2--8]}}=(1.1_{-0.3}^{+0.4})\times10^{-15}$\,\si{\flux}, $\chisq = 90.24$ with 100 dof). In both cases, there are no significant changes ($> 2\sigma$) to the dominant blackbody component, nor to $N_{\rm H}$. This excess had already been described in \citet{2019PASJ...71...17Y} and our results are compatible with their findings.

We then tested a three component model, akin to the one used for J1856, but the improvement of the best fit is not significant: we obtained $\chisq = 89.51$ with 98 dof. When compared to the best fitting model of a blackbody plus a power law, this model yields an F-test probability of 0.67. The best-fit parameters, expecially those of the power-law component, are poorly constrained: $kT_1^\infty=46.1\pm0.6$ eV, $R_1^\infty=4.6\pm0.2$; $kT_2^\infty=170\pm 50$ eV, $R_2^\infty=22\pm6$ m; $\Gamma=1.6_{-0.6}^{+2.0}$, $I_{\text{[2--8]}}=(2_{-1}^{+18}) \times 10^{-15}$\,\si{\flux}.
Therefore, we cannot confirm or dismiss the presence of two distinct hard components.

\section{Tables of the observations}
We report here the log of the observations used in this work: in Table~\ref{tab:observations_xmm} those from {\xmm} and in Table~\ref{tab:observations_nicer} the ones from {\nicer}.  Table~\ref{tab:observations_J04} reports the \xmm observations of RX\,J0420.0$-$5022.
\begin{table*}\centering
\caption{The \xmm observations used in this work. The filters are Thin1 (T1), Medium (M) and Thick (Tk), and for all the observations the EPIC-pn camera was operated in small window mode. The observations marked with a $\dagger$ were used only for the timing analysis and not for the spectral one.}
\label{tab:observations_xmm}
\begin{tabular}{l@{\hskip 1.5mm}c @{\hskip 1.5mm}c@{\hskip 1.5mm}c@{\hskip 3mm}r@{\hskip 2mm}c|@{\hskip 5mm}l@{\hskip 1.5mm}c @{\hskip 1.5mm}c@{\hskip 1.5mm}c@{\hskip 3mm}r@{\hskip 2mm}}\toprule

\# 	&	 Obs ID 	&	 Filter 	&	 Start date 	&	 Time (s) 	&	 \sartore	&	\# 	&	 Obs ID 	&	 Filter 	&	 Start date 	&	 Time (s) 	
\\\midrule																					
1	&	106260101	&	T1 	&	2002-04-08T16:22:29	&	58591	&	 A 	&	23	&	727760301	&	T1 	&	2014-09-18T11:03:44	&	78912	 \\ 
2	&	165971601	&	T1 	&	2004-09-24T01:42:13	&	33419	&	 B 	&	24	&	727760401	&	T1 	&	2015-03-12T11:46:22	&	74514	 \\ 
3	&	165971901	&	T1 	&	2005-03-23T08:34:42	&	35414	&	 C 	&	25	&	727760501	&	T1 	&	2015-10-03T15:56:32	&	81916	 \\ 
4	&	165972001	&	T1 	&	2005-09-24T07:58:13	&	35443	&	 D 	&	26	&	727760601	&	T1 	&	2016-03-11T21:51:55	&	76917	 \\ 
5	&	165972101	&	T1 	&	2006-03-26T15:40:29	&	70006	&	 E 	&	27	&	727761001	&	T1 	&	2016-09-23T00:14:46	&	70916	 \\ 
6	&	412600101	&	T1 	&	2006-10-24T00:31:14	&	73011	&	 F 	&	28	&	727761101	&	T1 	&	2017-03-15T06:45:19	&	69817	 \\ 
7	&	412600201	&	T1 	&	2007-03-14T20:50:01	&	69385	&	 G 	&	29	&	727761201	&	T1 	&	2017-09-16T17:42:56	&	74512	 \\ 
8	&	412600301	&	T1 	&	2007-10-04T05:48:55	&	70362	&	 I 	&	30	&	727761301	&	T1 	&	2018-04-10T12:18:35	&	69064	 \\ 
9	&	412600401	&	T1 	&	2008-03-13T18:47:14	&	74636	&	 J 	&	31$^\dagger$ 	&	791580101	&	Tk 	&	2016-04-16T17:08:12	&	18416	 \\ 
10	&	412600601	&	T1 	&	2008-10-05T01:00:58	&	68535	&	 K 	&	32$^\dagger$ 	&	791580201	&	M 	&	2016-04-16T22:33:13	&	9810	 \\ 
11	&	412600701	&	T1 	&	2009-03-19T21:30:04	&	68918	&	 L 	&	33	&	791580301	&	T1 	&	2016-04-17T01:34:53	&	6419	 \\ 
12	&	412600801	&	T1 	&	2009-10-07T12:06:44	&	81817	&	 M 	&	34$^\dagger$ 	&	791580401	&	Tk 	&	2016-04-17T03:39:55	&	19317	 \\ 
13	&	412600901	&	T1 	&	2010-03-22T02:48:57	&	73820	&	 N 	&	35$^\dagger$ 	&	791580501	&	M 	&	2016-04-17T09:19:56	&	8914	 \\ 
14	&	412601101	&	T1 	&	2010-09-28T23:09:10	&	69934	&	 O 	&	36	&	791580601	&	T1 	&	2016-04-17T12:06:36	&	16418	 \\ 
15	&	412601301	&	T1 	&	2011-03-14T00:45:15	&	83425	&	 P 	&	37	&	810840101	&	T1 	&	2018-10-19T06:59:24	&	69516	 \\ 
16	&	412601401	&	T1 	&	2012-04-13T07:14:06	&	77114	&	 - 	&	38	&	810840201	&	T1 	&	2019-04-12T16:31:11	&	77889	 \\ 
17$^\dagger$ 	&	412601501	&	T1 	&	2011-10-05T02:02:48	&	118098	&	 Q 	&	39	&	810841401	&	T1 	&	2019-09-18T22:00:20	&	68915	 \\ 
18	&	412602201	&	T1 	&	2013-03-14T08:26:38	&	73913	&	 - 	&	40	&	810841501	&	T1 	&	2020-03-31T23:16:37	&	72918	 \\ 
19	&	412602301	&	T1 	&	2012-09-20T11:25:04	&	80495	&	 - 	&	41	&	810841601	&	T1 	&	2020-09-15T14:48:30	&	73520	 \\ 
20	&	415180101	&	T1 	&	2007-03-25T05:36:47	&	40914	&	 H 	&	42	&	810841701	&	T1 	&	2021-04-01T08:36:25	&	70416	 \\ 
21	&	727760101	&	T1 	&	2013-09-14T14:20:54	&	71314	&	 - 	&	43	&	810841901	&	T1 	&	2021-10-11T12:59:23	&	71911	 \\ 
22	&	727760201	&	T1 	&	2014-03-26T06:00:22	&	73915	&	 - 	&	44	&	810842001	&	T1 	&	2022-04-03T00:05:28	&	73914	 \\ 

\bottomrule
\end{tabular}
\end{table*}

\begin{table*}\centering
\caption{The \nicer observations used in this work.}
\label{tab:observations_nicer}
\begin{tabular}{ccr|ccr|ccr}
\toprule
Obs ID	&	Start Date	&	Time (s)&Obs ID	&	Start Date	&	Time (s)&Obs ID	&	Start Date	&	Time (s)	\\
\midrule
1020520107	&		2019-02-16 03:02:37	&		3325	&	2614010109	&	2019-04-07 00:13:40	&		13177	&	2614010128	&	2019-05-22 16:20:37	&		1836	\\
1020520108	&		2019-02-17 00:42:11	&		4096	&	2614010110	&	2019-04-08 02:11:20	&		2111	&	2614010129	&	2019-05-23 03:03:22	&		961	\\
1020520109	&		2019-02-18 01:28:42	&		4195	&	2614010111	&	2019-04-08 23:43:46	&		5450	&	2614010130	&	2019-05-25 01:45:40	&		2566	\\
1020520110	&		2019-02-19 00:41:08	&		7081	&	2614010112	&	2019-04-10 00:37:20	&		4409	&	2614010131	&	2019-05-26 01:01:00	&		865	\\
1020520111	&		2019-02-20 01:28:20	&		649	&	2614010113	&	2019-04-11 02:50:00	&		5100	&	2614010132	&	2019-05-27 07:42:40	&		2931	\\
1020520112	&		2019-02-21 06:52:00	&		1615	&	2614010114	&	2019-04-13 12:09:00	&		629	&	2012110103	&	2019-05-27 15:50:00	&		696	\\
2020520102	&		2019-03-27 22:47:40	&		452	&	2614010115	&	2019-04-14 12:54:20	&		1668	&	2012120102	&	2019-05-27 17:23:00	&		338	\\
2020520103	&		2019-03-28 00:31:10	&		7466	&	2614010116	&	2019-04-16 17:14:20	&		1204	&	2614010133	&	2019-05-28 00:44:00	&		190	\\
2614010101	&		2019-03-30 20:42:19	&		735	&	2614010117	&	2019-04-19 10:08:00	&		1709	&	2012110104	&	2019-05-29 15:32:20	&		220	\\
2614010102	&		2019-03-31 02:39:20	&		6027	&	2614010118	&	2019-04-21 02:26:24	&		2507	&	2614010134	&	2019-06-12 01:37:40	&		368	\\
2614010103	&		2019-04-01 06:28:40	&		9325	&	2614010119	&	2019-04-22 00:15:40	&		8019	&	2614010135	&	2019-06-14 06:17:00	&		586	\\
2614010104	&		2019-04-02 01:19:38	&		5758	&	2614010120	&	2019-04-27 02:40:00	&		2125	&	2614010136	&	2019-06-15 00:51:20	&		12915	\\
2614010105	&		2019-04-03 01:32:18	&		5427	&	2614010121	&	2019-04-28 01:51:20	&		2594	&	2614010137	&	2019-06-16 04:43:00	&		1640	\\
2614010106	&		2019-04-04 01:09:41	&		897	&	2614010122	&	2019-04-29 00:43:00	&		1365	&	2614010138	&	2019-06-17 03:20:45	&		13534	\\
2614010107	&		2019-04-05 01:58:00	&		10883	&	2614010123	&	2019-05-05 11:01:21	&		95	&	2614010139	&	2019-06-18 01:10:40	&		5148	\\
2614010108	&		2019-04-06 00:37:00	&		16429	&	2614010127	&	2019-05-17 19:13:40	&		51	&							\\

\bottomrule
\end{tabular}
\end{table*}

\begin{table*}\centering
\caption{The \xmm observations of RX\,J0420.0$-$5022 used in this work. The science modes are Full frame (FF) and Small window (SW), and for all the observations the EPIC-pn camera was operated with thin optical filter.}
\label{tab:observations_J04}
\begin{tabular}{cc @{\hskip 1.5mm}c@{\hskip 1.5mm}c| cc @{\hskip 1.5mm}c@{\hskip 1.5mm}c}
\toprule

Obs ID & Science mode & Start date & Time (s) & Obs ID & Science mode & Start date & Time (s)\\\midrule
141750101	&	 FF  	&	 2002-12-30T03:38:44  	&	 17266	&	651470801	&	 SW  	&	 2010-10-02T23:05:56  	&	 8086 \\
141751001	&	 FF  	&	 2002-12-31T21:54:49  	&	 9879 	&	651470901	&	 SW  	&	 2010-10-03T19:17:37  	&	 9034 \\
141751101	&	 FF  	&	 2003-01-19T16:42:18  	&	 14212	&	651471001	&	 SW  	&	 2010-10-04T05:12:09  	&	 5285 \\
141751201	&	 FF  	&	 2003-07-25T21:21:51  	&	 17654	&	651471101	&	 SW  	&	 2010-10-06T22:57:07  	&	 5611 \\
651470201	&	 SW  	&	 2010-03-30T11:55:07  	&	 2899 	&	651471201	&	 SW  	&	 2010-11-26T09:28:48  	&	 3748 \\
651470301	&	 SW  	&	 2010-04-04T18:56:28  	&	 2570 	&	651471301	&	 SW  	&	 2011-01-13T22:23:20  	&	 3866 \\
651470401	&	 SW  	&	 2010-04-09T08:34:36  	&	 5442 	&	651471401	&	 SW  	&	 2011-03-31T20:15:41  	&	 4783 \\
651470501	&	 SW  	&	 2010-05-21T05:50:13  	&	 2433 	&	651471501	&	 SW  	&	 2011-04-11T07:13:22  	&	 3622 \\
651470601	&	 SW  	&	 2010-07-29T14:20:46  	&	 4363 	&	844140401	&	 FF  	&	 2019-05-22T07:52:39  	&	 16753 \\
651470701	&	 SW  	&	 2010-09-21T08:40:34  	&	 6764 &&&\\

\bottomrule
\end{tabular}
\end{table*}

\bsp	
\label{lastpage}
\end{document}